\documentclass[aps,twocolumn,  amsmath,amssymb,
 aps,prc,superscriptaddress]{revtex4-2}
\usepackage[utf8]{inputenc}
\usepackage[english]{babel}
\usepackage{listings}
\usepackage{algorithm}
\usepackage{algorithmicx}
\usepackage{algcompatible}
\usepackage{adjustbox}
\usepackage{graphicx}
\usepackage{color}
\usepackage[normalem]{ulem}

\usepackage{tikz}
\usepackage{caption}
\usepackage{amsmath}
\usepackage[caption=false,font=footnotesize]{subfig}
\usepackage{placeins}
\usepackage{hyperref}
\usepackage{url}
\usepackage{lipsum}
\definecolor{gray97}{gray}{.97}
\definecolor{gray75}{gray}{.75}
\definecolor{gray45}{gray}{.45}

\def\BibTeX{{\rm B\kern-.05em{\sc i\kern-.025em b}\kern-.08em
    T\kern-.1667em\lower.7ex\hbox{E}\kern-.125emX}}

\graphicspath{{.}{./Figures/}}

\begin{document}

\title{Utilization of Event Shape in Search of the Chiral Magnetic Effect in Heavy-ion Collisions}

\author{Ryan Milton}
\email{rmilton@ucla.edu}
\affiliation{Department of Physics and Astronomy, University of
  California, Los Angeles, California 90095, USA}

\author{Gang Wang}
\email{gwang@physics.ucla.edu}
\affiliation{Department of Physics and Astronomy, University of
  California, Los Angeles, California 90095, USA} 

\author{Maria Sergeeva}
\affiliation{Department of Physics and Astronomy,
  University of California, Los Angeles, California 90095, USA}
  
\author{Shuzhe~Shi}
\affiliation{Department of Physics, McGill University, 3600 University Street, Montreal, QC, H3A 2T8, Canada}

\author{Jinfeng~Liao}
\affiliation{
Physics Department and Center for Exploration of Energy and Matter, Indiana University, 2401 N Milo B. Sampson Lane, Bloomington, IN 47408, USA}

\author{Huan Zhong Huang} \affiliation{Department of Physics and
Astronomy, University of California, Los Angeles, California 90095,
  USA}
  \affiliation{Key Laboratory of Nuclear Physics and Ion-beam
  Application (MOE), and Institute of Modern Physics, Fudan
  University, Shanghai-200433, People’s Republic of China}

\begin{abstract}
{The search for the chiral magnetic effect (CME) has been a subject of great interest in the field of high-energy heavy-ion collision physics, and various observables have been proposed to probe the CME. Experimental observables are often contaminated with background contributions arising from collective motions (specifically elliptic flow) of the collision system. We present a method study of event-shape engineering (ESE) that projects the CME-sensitive $\gamma_{112}$ correlator and its variations ($\gamma_{132}$ and $\gamma_{123}$) to a class of events with minimal flow. We discuss
the realization of the zero-flow mode, the sensitivity on the CME signal, and the corresponding statistical significance for Au+Au, Ru+Ru, and Zr+Zr collisions at $\sqrt{s_{\rm NN}} = 200$ GeV with a multiphase transport (AMPT) model,
as well as a new event generator, Event-By-Event Anomalous-Viscous Fluid Dynamics (EBE-AVFD).}
\end{abstract}


\maketitle

\section{Introduction}

A major goal of the experiments on high-energy heavy-ion collisions is to produce a deconfined nuclear matter, known as the Quark-Gluon Plasma (QGP), and to study its properties.
The creation of a QGP provides a test to the topological sector of
quantum chromodynamics (QCD),
the fundamental theory of strong interactions. 
According to QCD, quarks in a QGP could obtain a chirality imbalance via the chiral anomaly~\cite{ChiralAnomality1,ChiralAnomality2},
forming local 
domains with finite chiral chemical potentials ($\mu_5$)~\cite{Kharzeev_PLB2006,Kharzeev_NPA2008,Kharzeev_NPA2007,Kharzeev_PLB2002,Yin_PRL2015,Kharzeev_PRL2010}. These chiral quarks could manifest an electric current along the direction of the strong magnetic field ($\overrightarrow{B} \sim 10^{14}$ T) generated by the incident protons in the heavy-ion collisions: $\overrightarrow{J_e} \propto \mu_5\overrightarrow{B}$,
which is theorized as the chiral magnetic
effect (CME)~\cite{Kharzeev_PLB2006,Kharzeev_NPA2008}. Some recent reviews on the CME are available in Refs.~\cite{Review1,Review2,Review3,Kharzeev:2020jxw}.

\begin{figure}[h]
\centering
\includegraphics[width=0.45\textwidth]{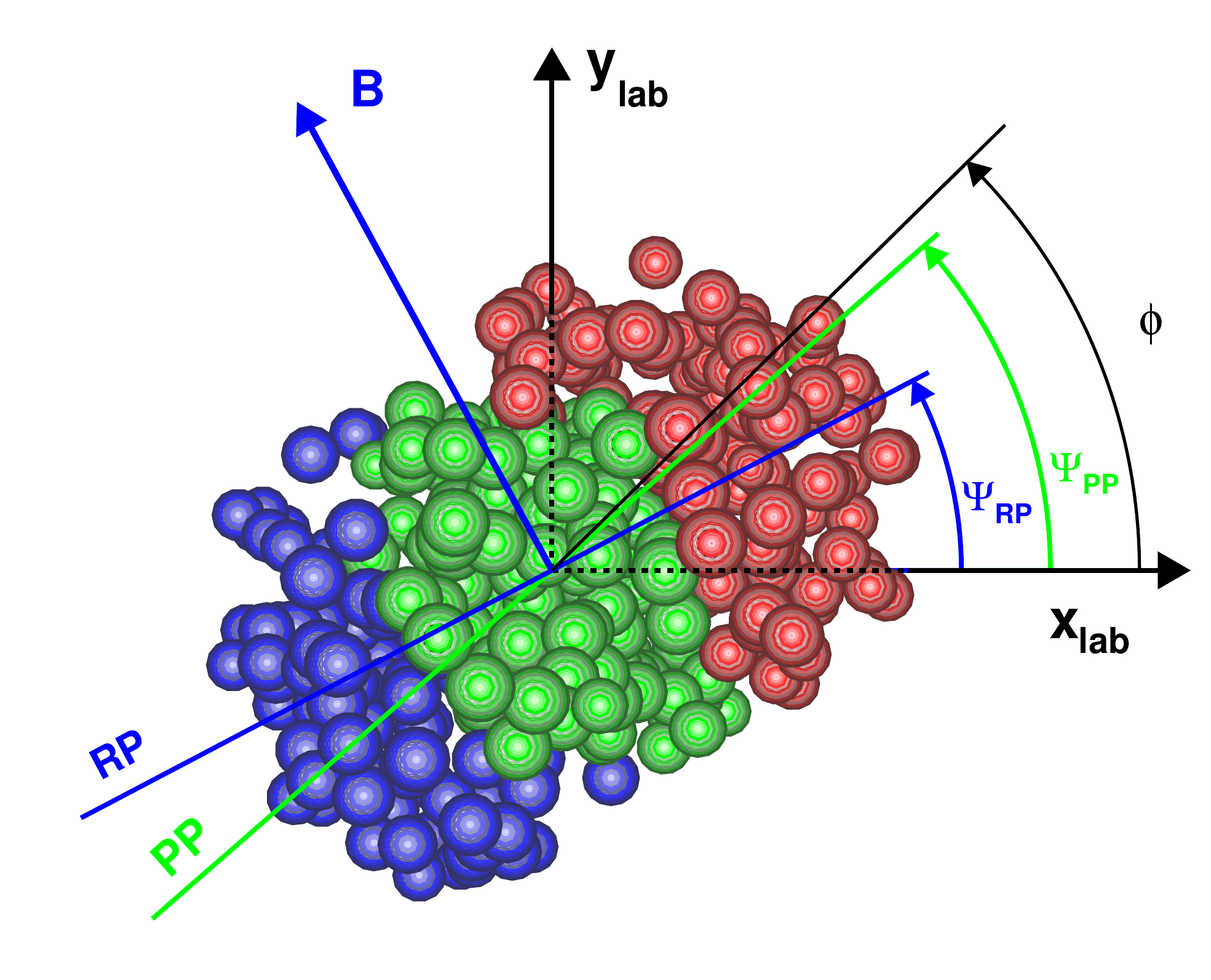}
\caption{ Schematic diagram of the transverse
plane for a two-nucleus collision, with the left one emerging
from and the right one going into the page. Particles are
produced in the overlap region (green-colored participating nucleons). The
azimuthal angles of the reaction plane ($\Psi_{\rm RP}$), the participant plane ($\Psi_{\rm PP}$), and a produced particle
($\phi$) are depicted here.}
\label{fig:RP}
\end{figure}

On average, $\overrightarrow{B}$ is expected to be perpendicular to the reaction plane (RP), which is spanned by the impact
parameter and the beam momenta of a collision. The CME will then give rise to an electric charge separation across the RP.
In the study of the CME-induced charge separation as well as other collective motions in the QGP, the azimuthal distribution of produced particles is often expressed with the Fourier expansion for given transverse momentum ($p_T$) and pseudorapidity ($\eta$) in an event:
\begin{eqnarray}
\begin{aligned}
\frac{dN_{\alpha}}{d\phi^*} &\approx \frac{N_\alpha}{2\pi} [1 + 2v_{1,\alpha}\cos(\phi^*) + 2v_{2,\alpha}\cos(2\phi^*) \\
& +  2v_{3,\alpha}\cos(3\phi^*) + ... +2a_{1,\alpha}\sin(\phi^*) + ... \ ] , 
\label{eq:fourier}
\end{aligned}
\end{eqnarray}
where $\phi^* = \phi - \Psi_{\rm RP}$, and $\phi$ and $\Psi_{\rm RP}$ are the azimuthal angles of a particle and the RP, respectively.
The subscript $\alpha$ ($+$ or $-$) denotes the particle's charge sign.
Traditionally, the coefficients $v_1$, $v_2$, and $v_3$ are called ``directed flow", ``elliptic flow", and ``triangular flow", respectively. 
In the scenario of fluid evolution, these $v_n$ coefficients 
reflect the hydrodynamic response of the QGP to the initial collision geometry and to its fluctuations~\cite{HYDRO_review}. Figure~\ref{fig:RP} sketches the transverse plane, perpendicular to the beam direction (the $z_{\rm lab}$ axis), in an off-center heavy-ion collision.
In practice, an event plane obtained from the collective motion of detected particles is used instead of the true RP. For simplicity, we still use the RP notation in the following discussions, and RP could represent a specific event plane.

The $a_1$ coefficient (with $a_{1,+}\approx-a_{1,-}$ in a charge-symmetric system) quantifies the CME-induced charge separation. At first glance, it seems $a_1$ can be measured via $\langle \sin(\phi^*) \rangle$, averaged over particles in each event and then over all events, to probe the CME. However, $\mu_{5}$ flips sign on an event-by-event basis with equal probability, forcing $\langle\sin(\phi^*) \rangle$ to be zero. Therefore, several observables in search of the CME have been designed to measure $a_{1,\pm}$ fluctuations across the RP,
such as the $\gamma$ correlators~\cite{Sergei2004}, the $R$ correlators~\cite{PHENIX2,RCorr-2018}, and the signed balance functions~\cite{SBF-Aihong-2020,Yufu-2020}. It has been demonstrated that all these methods carry essentially the same physical message with similar sensitivities to the CME signal and backgrounds~\cite{CME_methods}. In this paper we focus on the $\gamma_{112}$ correlator~\cite{Sergei2004},
\begin{eqnarray}
\gamma_{112} &\equiv&  \langle \cos(\phi_\alpha + \phi_\beta -2{\rm \Psi_{RP}}) \rangle  \\
&=& \langle\cos(\phi^*_{\alpha})\cos(\phi^*_{\beta}) -
\sin(\phi^*_{\alpha})\sin(\phi^*_{\beta})\rangle \nonumber \\
&=& (\langle v_{1,\alpha}v_{1,\beta}\rangle + B_{\rm IN}) -(\langle a_{1,\alpha}a_{1,\beta}\rangle + B_{\rm OUT}),\nonumber \label{eq:ThreePoint}
\end{eqnarray}
where the averaging is done over all pairs of particles $\alpha$ and $\beta$ in each event and over all events.
$\langle a_{1,\alpha}a_{1,\beta}\rangle$
is the main target of the CME search, whereas $\langle v_{1,\alpha}v_{1,\beta}\rangle$ is expected to be charge-independent and unrelated to the electromagnetic field in symmetric A+A collisions.
$B_{\rm IN}$ and $B_{\rm OUT}$ represent other
possible in-plane and out-of-plane background correlations, respectively.

The difference between the $\gamma_{112}$ correlators for opposite-sign pairs and same-sign pairs, 
\begin{equation}
\Delta \gamma_{112} \equiv \gamma^{\rm OS}_{112} - \gamma^{\rm SS}_{112} \label{eq:Dg112}, 
\end{equation}
adds up the CME contributions, and cancels out the charge-independent backgrounds.
However, there are still some residual backgrounds, such as those originating from decays of flowing resonances~\cite{Wang:2009kd,Schlichting:2010qia},
transverse momentum conservation (TMC)~\cite{Pratt:2010zn,Bzdak:2012ia}, 
and local charge conservation (LCC)~\cite{Schlichting:2010qia}.
In general, these background mechanisms can be regarded as the coupling between elliptic flow ($v_2$) and the two-particle correlation,
\begin{equation}
\delta \equiv \langle \cos(\phi_\alpha -\phi_\beta) \rangle.
\label{eq:delta}
\end{equation}
The goal of an event-shape-engineering (ESE) approach is to project the $\Delta\gamma_{112}$ measurements to a class of events with zero $v_2$ (or zero $v_2 \Delta\delta$) to remove the non-CME background. Note that $\Delta\delta \equiv \delta^{\rm OS} - \delta^{\rm SS}$. We will adopt the technique in Ref.~\cite{ESE1}
to make an event-shape selection such that the particles of interest (POI) form an almost spherical sub-event, bearing close-to-zero anisotropic flow.

Besides $\gamma_{112}$ that contains both the CME signal and the background, other correlators have also been proposed that are overwhelmed by the background contributions, such as~\cite{intro_15a,intro_16,Subikash} 
\begin{equation}
\gamma_{132}\equiv\langle \cos{(\phi_{\alpha}-3\phi_{\beta}+2\Psi_{2})} \rangle / Res\{\Psi_{2}\} 
\end{equation} 
and
\begin{equation}
\gamma_{123}\equiv\langle \cos{(\phi_{\alpha}+2\phi_{\beta}-3\Psi_{3})} \rangle / Res\{\Psi_{3}\} ,
\end{equation}
where $\Psi_{2}$ and $\Psi_{3}$ denote the $2^{\rm nd}$-order and $3^{\rm rd}$-order event planes, respectively. Here the $n^{\rm th}$-order event plane is estimated with the $v_n$ information of detected particles.
The measurements of $\gamma_{112}$, $\gamma_{132}$, and $\gamma_{123}$ with respect to specific event planes need to be corrected with the corresponding event plane resolutions ($Res\{\Psi_{2}\} $ and $Res\{\Psi_{3}\}$).

In Ref.~\cite{ESE1}, an ESE recipe has been applied to Au+Au events simulated by a multiphase transport (AMPT)~\cite{ampt_1} model, in a pure-background scenario, to demonstrate the disappearance of the background in $\Delta\gamma_{112}$. 
In this article, we not only extend the AMPT study to $\gamma_{132}$ and $\gamma_{123}$, but also employ a new event generator, Event-By-Event Anomalous-Viscous Fluid Dynamics (EBE-AVFD)~\cite{Shi:2017cpu,Jiang:2016wve,Shi:2019wzi}, which implements the CME signal on top of the flow-related background.
Thus, we explore the sensitivity of the ESE approach to both the realistic backgrounds and the CME signal.

The STAR experiment at RHIC has collected a large data sample of isobar collisions, namely $^{96}_{44}$Ru + $^{96}_{44}$Ru and $^{96}_{40}$Zr + $^{96}_{40}$Zr at $\sqrt{s_{\rm NN}}=200$ GeV, in search of the possible difference in the CME-induced charge separation in these collisions. The two isobaric systems have the same number of nucleons and hence similar amounts of elliptic flow, but different numbers of protons, roughly leading to a 15\% difference in the magnetic field squared and in turn, a similar magnitude difference in the CME signal~\cite{UU_theory,isobar1,isobar2}. The sensitivities of several observables to the CME in these isobar collisions have been investigated using the EBE-AVFD simulations~\cite{CME_methods},
and we will examine whether the ESE approach has any advantage over the ensemble average in $\Delta\gamma_{112}$.

In Sec. II, we
shall review the ESE method, including how to select the event shape handle, how to suppress the flow-related background, and how to restore the ensemble average of the CME signal. Then the simulation studies with AMPT and EBE-AVFD will be presented in Sec.~III. Section IV gives the summary and the outlook.

\section{Methodology}

There are three key components to a successful ESE approach. First, a direct handle on the event shape should be able to reflect the initial geometrical configuration
for the POIs on an event-by-event basis. Second, the flow-induced backgrounds have to vanish or at least be greatly suppressed at the zero-flow mode. Third, the true CME signal ought to be restored to the ensemble average value from the zero-flow mode. We will elaborate on these aspects in the following discussions.

The ``standard" ESE procedure is to keep the following three types of particles independent of each other in an event:
(A) the particles that define the event shape, (B) the POIs ($\alpha,\beta$), and (C) the particles that reconstruct the event plane ($\Psi_{\rm EP}$). In other words, they should come from three distinct sub-events. Conventionally, the event shape is controlled by the magnitude of the flow vector of sub-event A, $\overrightarrow{q_n}^{\rm A} = (q_{n,x}^{\rm A},q_{n,y}^{\rm A})$:
\begin{eqnarray}
q_{n,x}^{\rm A} &=& \frac{1}{\sqrt{N}} \sum_i^N \cos(n\phi_i^{\rm A}), \label{qx}  \\
q_{n,y}^{\rm A} &=& \frac{1}{\sqrt{N}} \sum_i^N \sin(n\phi_i^{\rm A}), \label{qy}
\end{eqnarray}
which is related to $v_n$ with some extra statistical fluctuations.
For events in each $q_n^{\rm A}$ class, $v_n^{\rm B}$ and various $\Delta\gamma^{\rm B}$ correlators are calculated for the POIs in sub-event B, with the event plane estimated from sub-event C.
Then collected from all the $q_n^{\rm A}$ classes, $\Delta\gamma_{112}^{\rm B}$ or $\Delta\gamma_{132}^{\rm B}$ ($\Delta\gamma_{123}^{\rm B}$) is plotted as a function of $v_2^{\rm B}$ ($v_3^{\rm B}$), and the extrapolation to zero $v_n^{\rm B}$ gives the $\Delta\gamma^{\rm B}$ results at the zero-flow mode. Although $q_n^{\rm A}$ and $q_n^{\rm B}$ are linearly correlated on average,
there is a spread between them on an event-by-event basis, arising from statistical fluctuations. Consequently, even the lowest $q_n^{\rm A}$ bin (close to zero) corresponds to a positive and sizable $v_n^{\rm B}$. Systematic uncertainties 
and model dependence have to be introduced when $\Delta\gamma^{\rm B}$ is extrapolated over a wide unmeasured $v_n^{\rm B}$ region. Hence, this ``standard" approach only provides an indirect handle on the event shape for the POIs.

In order to avoid the long extrapolation in $v_n^{\rm B}$, we follow the recipe in Ref.~\cite{ESE1} to forsake the independence between sub-events A and B.
For simplicity, we will omit the superscript ``A" or ``B" in $q_n$, $v_n$, or $\Delta\gamma$ in the following discussions.
The merging of sub-events A and B not only reduces the statistical uncertainty for both $q_n$ and the POIs, but also makes the lowest $q_n$ bin naturally correspond to a very small $v_n$ value. Therefore the extrapolation to the zero-flow mode is technically much more reliable. The caveat is that only the ``apparent" flow for the POIs is under control.
It is still possible that although a resonance parent has a finite $v_n$ value, its decay daughters give zero contribution to $q_n$~\cite{Wang:2016iov}. In this case, even at zero $q_n$ or $v_n$, there exists a finite non-CME contribution in the $\Delta\gamma$ correlators. If that happens, we cannot completely remove the flow-related background, but only suppress it to a large extent. Such residual backgrounds will be investigated with realistic models in the next section.

In the upcoming model studies, we select mid-rapidity particles with $|\eta|<1$ and $0.15 < p_T < 2$ GeV/$c$ to form sub-event A(B).
For sub-event C, we will exploit both the RP and the participant plane (PP), defined by the initial density distribution of the participating nucleons. Measurements with respect to $\Psi_{\rm RP}$ and $\Psi_{\rm PP}$ bear different sensitivities to the CME signal and the flow-induced background~\cite{Shi:2019wzi}. In real-data analyses, $\Psi_{\rm RP}$ is usually approximated by the spectator plane, determined by a sideward deflection of spectator nucleons. Since $\Psi_{\rm RP}$ is preset in the simulation framework, we will directly use the known $\Psi_{\rm RP}$ in this method study.
On the other hand, particles (still with $0.15 < p_T < 2$ GeV/$c$) with $1.5 <\eta<5$ and $-5 <\eta<-1.5$ will be employed to reconstruct two separate flow vectors, and the corresponding azimuthal angles ($\Psi_{\rm EP1}$ and $\Psi_{\rm EP2}$) will serve as an
estimate of $\Psi_{\rm PP}$.
For events in each $q_n$ bin, the square root of $\langle \cos[n(\Psi_{\rm EP1} - \Psi_{\rm EP2})]\rangle$ renders the pertinent sub-event plane resolution. In the following discussions, whenever we present the results measured with respect to $\Psi_{\rm PP}$, the resolution effect will have been corrected beforehand, and will not be mentioned anymore.

Although both $q_n$ and $q_n^2$ well characterize the event shape, we prefer $q_n^2$,
because its distribution peaks around zero~\cite{ESE1}, yielding a more reliable projection of $\Delta\gamma$ to the zero-flow mode. For events within each $q_2^2$ or $q_3^2$ interval, we perform a set of measurements of ($v_2$, $\Delta\delta$, $\Delta\gamma_{112}$, $\Delta\gamma_{132}$) or ($v_3$, $\Delta\delta$, $\Delta\gamma_{123}$), respectively. After these observables are measured over the whole $q_n^2$ range under study, the zero-flow mode is achieved by projecting the $\Delta\gamma$ correlators to $q_n^2 = 0$, or alternatively, to $v_n=0$ or $v_n \Delta\delta = 0$. 
The latter dependence arises naturally from specific background mechanisms, and could provide valuable insight.

In the presence of a CME-induced charge separation ($\Delta\gamma_{112}^{\rm CME}$), the intercept of $\Delta\gamma_{112}|_{q_2^2=0}$ is expected to be positively finite. This intercept, however, is not equal to the ensemble average of $\Delta\gamma_{112}^{\rm CME}$, because there is an intrinsic relation between the event-by-event quantities of $\Delta\gamma_{112}^{\rm CME}$ and $v_2$ (and hence $q_2^2$, when both are obtained from the same sub-event). Following the derivation in Ref.~\cite{ESE1}, we need to multiply $\Delta\gamma_{112}|_{q_2^2=0}$ by a factor of $(1-2v_2)$ to restore the ensemble average of $\Delta\gamma_{112}^{\rm CME}$.

\section{Results}

We shall first use the pure-background AMPT model to test the behaviors of $\Delta\gamma_{112}$ and $\Delta\gamma_{132}$ at the zero-flow mode with both the RP and the PP, as well as that of $\Delta\gamma_{123}$ with respect to the PP. Note that
$v_3$ and $\Psi_3$ originate from the statistical fluctuation of the initial geometry of the participating zone, and the measurements of $v_3$ and $\gamma_{123}$ naturally yield zero with respect to the RP.
Next, the responses of the $\gamma$ correlators to the CME signal and the background will be further explored with the ESE approach using events simulated by the EBE-AVFD model.

\subsection{AMPT}

AMPT is a hybrid transport event simulator that divides a heavy-ion collision into four stages: the initial conditions, the partonic evolution, the hadronization,
and the hadronic interactions. For the initial conditions, AMPT adopts the
spatial and momentum distributions of minijet partons and excited soft strings from the Heavy Ion Jet Interaction Generator (HIJING) \cite{ampt_2}.
Then Zhang's parton cascade (ZPC) \cite{ampt_3}
is deployed to run the partonic evolution, characterized
by two-body parton-parton elastic scattering.
Towards the end of the partonic evolution,
the spatial quark coalescence is implemented to attain the quark-hadron phase transition in the string melting (SM) version of AMPT.
Finally, the hadronic interactions are modelled by A Relativistic Transport calculations (ART)~\cite{ampt_4}.

The SM version of AMPT
reasonably well reproduces particle spectra and elliptic
flow in Au+Au collisions at 200 GeV and Pb+Pb collisions at 2.76 TeV~\cite{ampt_5}.
In this study, the SM v2.25t4cu of AMPT has been used to simulate $2.4\times10^7$ events of 0--80\% Au+Au collisions at $\sqrt{s_{\rm NN}}=$ 200 GeV.
This version conserves electric charge, which is particularly important for the CME-related analyses.
The model parameters are set in the same way as in Ref.~\cite{Subikash}. Only $\pi^\pm$, $K^\pm$, $p$ and $\bar{p}$ are included in the following simulations.

\begin{figure}[tb]
\centering
\includegraphics[width=0.42\textwidth]{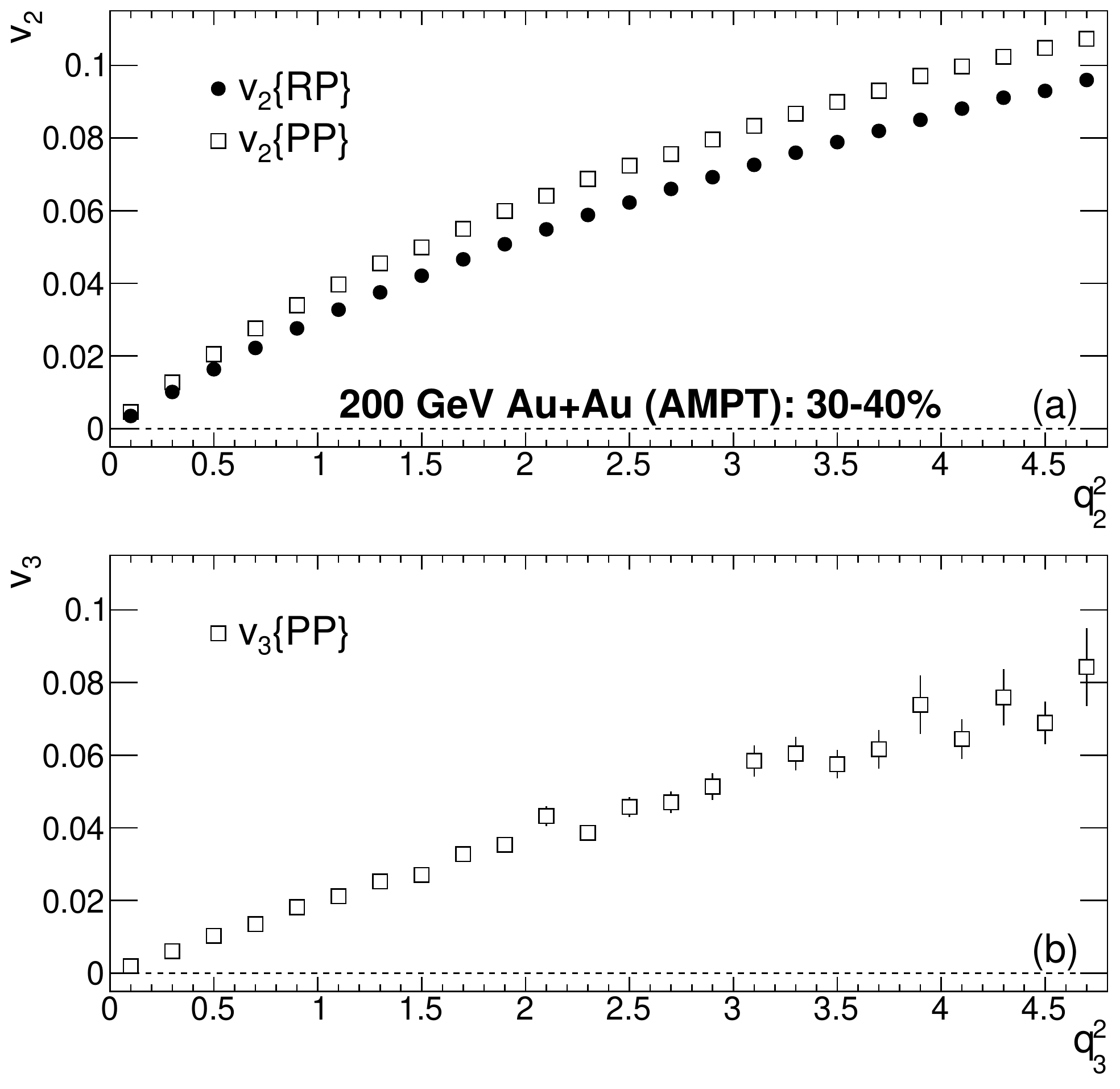}
\caption{AMPT simulations of $v_2$ vs $q^2_2$ (a) and $v_3$ vs $q^2_3$ (b) for 30--40\% Au+Au collisions at $\sqrt{s_{\rm NN}}=200$ GeV.}
\label{fig:AMPT_vn_q}
\end{figure}

\begin{figure}[h]
\centering
\includegraphics[width=0.43\textwidth]{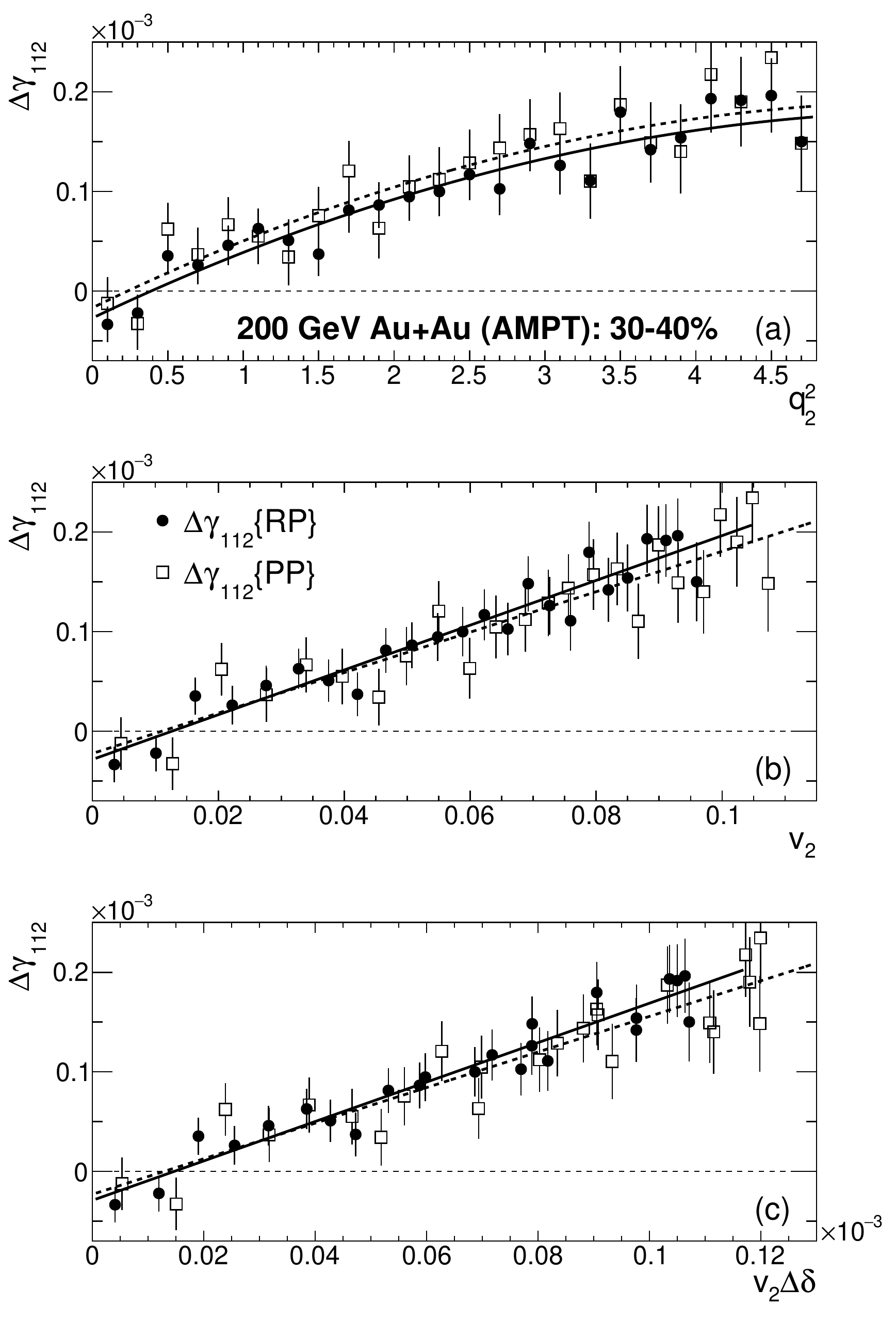}
\caption{AMPT calculations of $\Delta\gamma_{112}$ as a function of $q^2_2$ (a), $v_2$ (b), and $v_2\Delta\delta$ (c) for 30--40\% Au+Au collisions at $\sqrt{s_{\rm NN}}=200$ GeV. The results are fit with second-order polynomial functions in panel (a), and with linear functions in panels (b) and (c).
}
\label{fig:AMPT_Dg112_q}
\end{figure} 

\begin{figure}[tb]
\centering
\includegraphics[width=0.43\textwidth]{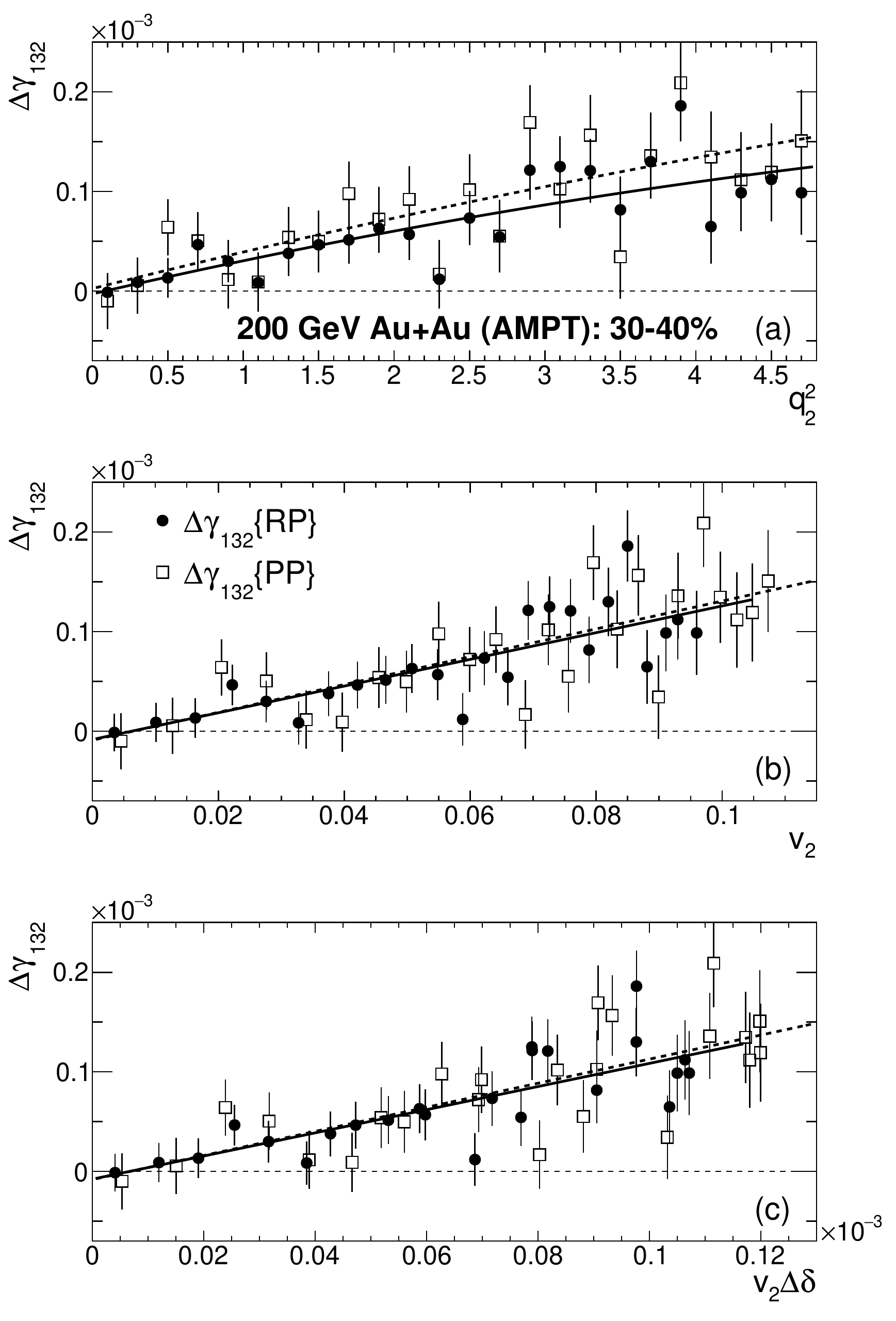}
\caption{AMPT simulations of $\Delta\gamma_{132}$ as a function of $q^2_2$ (a), $v_2$ (b), and $v_2\Delta\delta$ (c) for 30--40\% Au+Au collisions at $\sqrt{s_{\rm NN}}=200$ GeV. The results are fit with second-order polynomial functions in panel (a), and with linear functions in panels (b) and (c).}
\label{fig:AMPT_Dg132_q}
\end{figure} 

\begin{figure}[tb]
\centering
\includegraphics[width=0.43\textwidth]{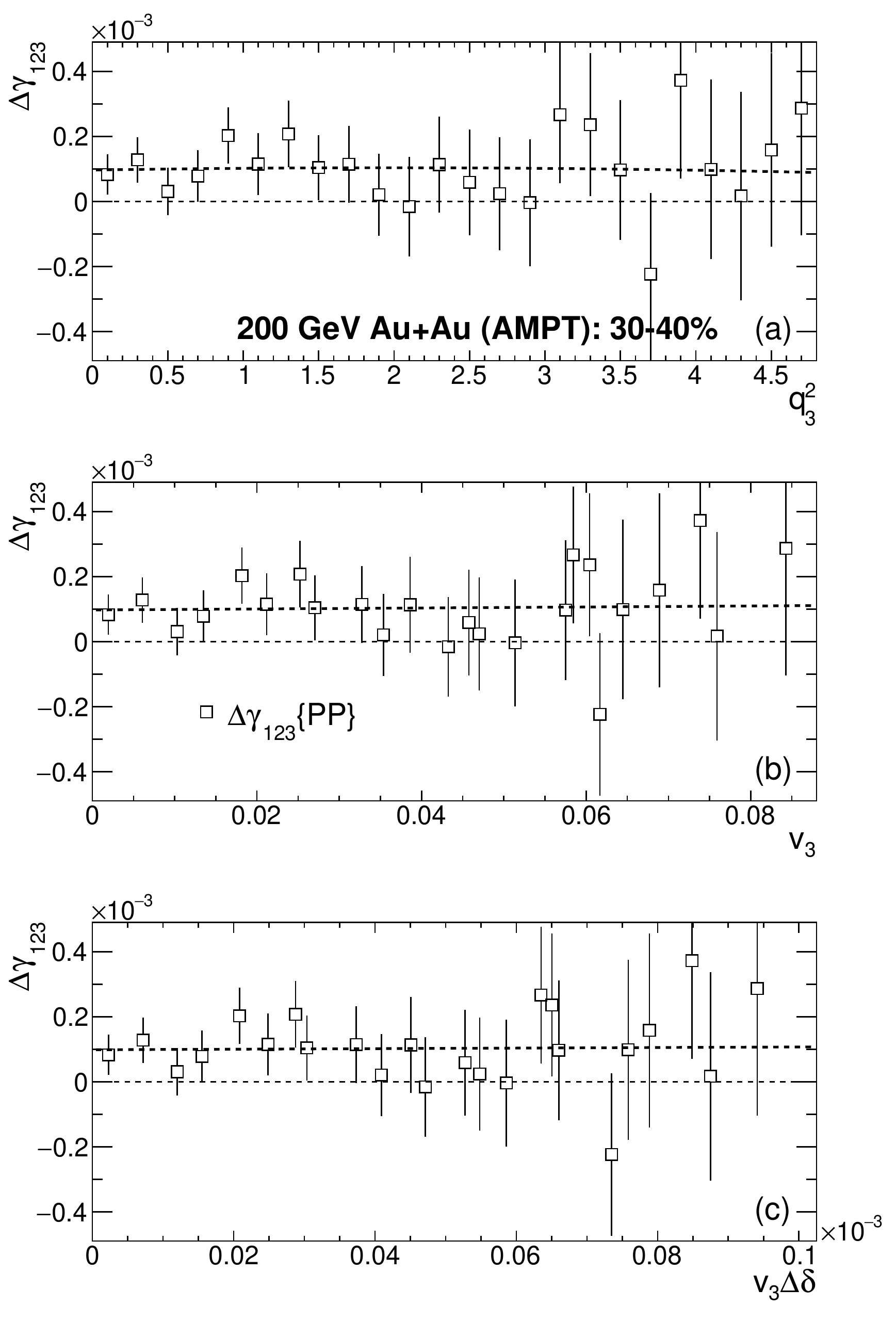}
\caption{AMPT calculations of $\Delta\gamma_{123}$ as a function of $q^2_3$ (a), $v_3$ (b), and $v_3\Delta\delta$ (c) for 30--40\% Au+Au collisions at $\sqrt{s_{\rm NN}}=200$ GeV. The results are fit with a second-order polynomial function in panel (a), and with linear functions in panels (b) and (c).}
\label{fig:AMPT_Dg123_q}
\end{figure}

Figure~\ref{fig:AMPT_vn_q} demonstrates the effectiveness of $q_2^2$ and $q_3^2$ in characterizing the event shape with AMPT calculations of $v_2(q_2^2)$ (a) and $v_3(q_3^2)$ (b), respectively, for the 30--40\% centrality interval in Au+Au collisions at $\sqrt{s_{\rm NN}}=200$ GeV. Although the $v_2$ values with respect to the RP and the PP are different because of the fluctuation of the initial nucleons,
 they both
approach zero at $q_2^2 = 0$.
In a similar way, the zero-$v_3$ mode is also realized at vanishing $q_3^2$. The $q_2^2$ and 
$q_3^2$ ranges under study have covered 86.2\% and 98.5\% of the whole event sample, respectively.

\begin{figure}[tb]
\centering
\includegraphics[width=0.42\textwidth]{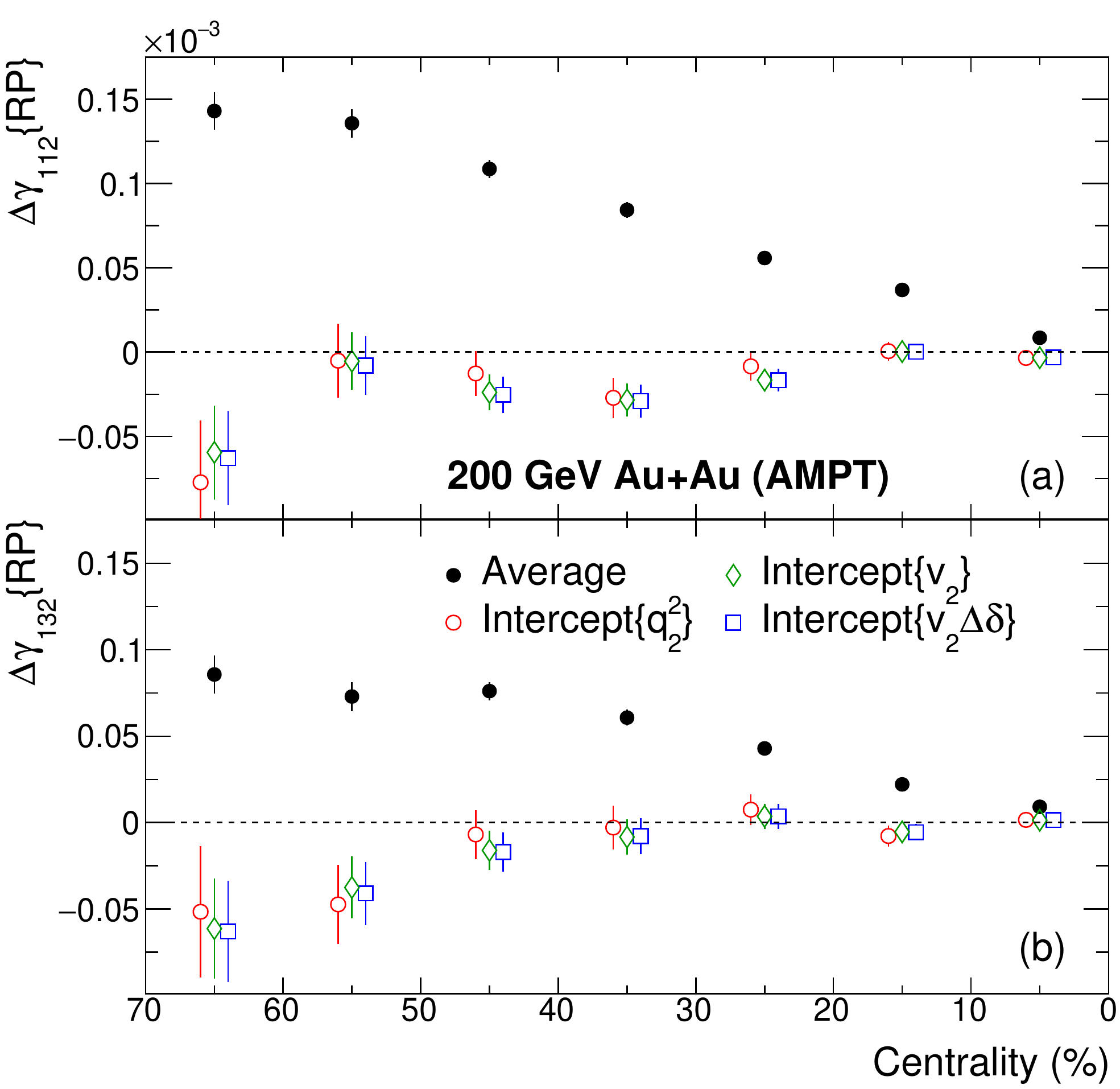}
\caption{Centrality dependence of $\Delta\gamma_{112}\{\rm RP\}$ (a) and $\Delta\gamma_{132}\{\rm RP\}$ (b) at the zero-flow mode for AMPT events of Au+Au collisions at $\sqrt{s_{\rm NN}}=200$ GeV.
The open markers represent the fit intercepts via different variables: $q_2^2$, $v_2$, and $v_2\Delta\delta$. In comparison, the ensemble averages are also drawn with the solid markers. }
\label{fig:AMPT_Dg_cen_RP}
\end{figure}

Figures~\ref{fig:AMPT_Dg112_q} and~\ref{fig:AMPT_Dg132_q} illustrate how the zero-flow mode is accomplished for $\Delta\gamma_{112}$ and $\Delta\gamma_{132}$, respectively, via $q^2_2$ (a), $v_2$ (b), and $v_2\Delta\delta$ (c) with AMPT events of 30--40\% Au+Au collisions at $\sqrt{s_{\rm NN}}=200$ GeV.
Each panel contains results with respect to both the RP and the PP, which are fit with second-order polynomial functions in panel (a), and with linear functions in panels (b) and (c).
The solid (dashed) lines represent the fit functions to the results with respect to the RP (PP). In panel (a), the solid and dashed lines 
are significantly different, since the flow-related background is proportional to $v_2$, and $v_2\{\rm RP\}$ is different from $v_2\{\rm PP\}$, as shown in Fig.~\ref{fig:AMPT_vn_q}(a). In panels (b) and (c), where $v_2$ explicitly appears on the horizontal axis, the difference between the solid and dashed lines is suppressed. The vertical intercepts of all the fit functions are close to zero, but slightly negative, indicating a potential over-correction of the flow-related background.
The centrality dependence of these intercepts will be discussed later in Figure~\ref{fig:AMPT_Dg_cen_RP}.

Figure~\ref{fig:AMPT_Dg123_q} exhibits $\Delta\gamma_{123}$ with respect to the PP vs $q_3^2$ (a), 
$v_3$ (b), and $v_3\Delta\delta$ (c), calculated with AMPT events of 30--40\% Au+Au collisions at $\sqrt{s_{\rm NN}}=200$ GeV.
The results are fit with a second-order polynomial function in panel (a), and with linear functions in panels (b) and (c). Unlike the cases of $\Delta\gamma_{112}$ and $\Delta\gamma_{112}$, $\Delta\gamma_{123}$ does not diminish at the zero-flow mode, but stays rather constant as a function of either $q_3^2$, $v_3$, or $v_3\Delta\delta$.
This implies that the formation of the finite $\Delta\gamma_{123}$ arises from a different mechanism than either $\Delta\gamma_{132}$ or the flow-induced background in $\Delta\gamma_{112}$.

\begin{figure}[h]
\centering
\includegraphics[width=0.39\textwidth]{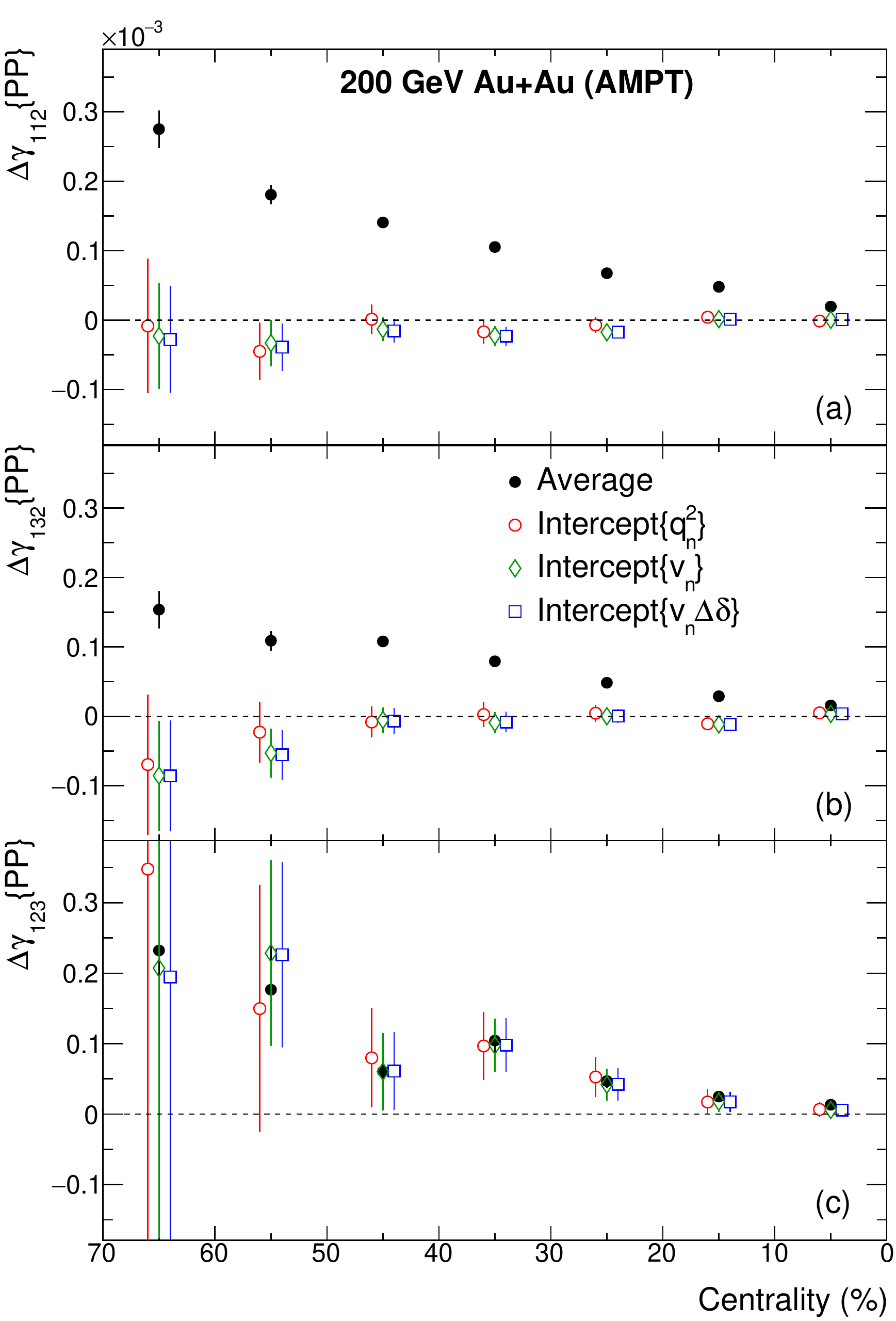}
\caption{Centrality dependence of $\Delta\gamma_{112}\{\rm PP\}$ (a), $\Delta\gamma_{132}\{\rm PP\}$ (b), and $\Delta\gamma_{123}\{\rm PP\}$ (c) at the zero-flow mode for AMPT events of Au+Au collisions at $\sqrt{s_{\rm NN}}=200$ GeV.
The open markers represent the fit intercepts via different variables: $q_n^2$, $v_n$, and $v_n\Delta\delta$. In comparison, the ensemble averages are also drawn with the solid markers.}
\label{fig:AMPT_Dg_cen_PP}
\end{figure} 


We have applied the same analysis procedure as in Figs.~\ref{fig:AMPT_Dg112_q},~\ref{fig:AMPT_Dg132_q}, and~\ref{fig:AMPT_Dg123_q} to different centrality classes. Figure~\ref{fig:AMPT_Dg_cen_RP} shows the centrality dependence of $\Delta\gamma_{112}$ (a) and $\Delta\gamma_{132}$ (b) with respect to the RP at the zero-flow mode for AMPT events of Au+Au collisions at $\sqrt{s_{\rm NN}}=200$ GeV.
The open markers represent the fit intercepts via different variables: $q_2^2$, $v_2$, and $v_2\Delta\delta$. At each centrality interval, the three intercepts are consistent with each other, and are consistent with or lower than zero. Therefore, the zero-flow projection in AMPT events demonstrates similar efficacy in removing the flow-related background in 
$\Delta\gamma_{112}$ and $\Delta\gamma_{132}$. 
In comparison, the ensemble averages are also drawn with the solid markers.
Using the ensemble averages
as a reference baseline, 
we illustrate how much background contributions have been suppressed with the ESE technique, and also visualize the potential over-subtraction of background in some centrality ranges. 

Figure~\ref{fig:AMPT_Dg_cen_PP} displays the centrality dependence of $\Delta\gamma_{112}$ (a), $\Delta\gamma_{132}$ (b), and $\Delta\gamma_{123}$ (c) with respect to the PP at the zero-flow mode for AMPT events of Au+Au collisions at $\sqrt{s_{\rm NN}}=200$ GeV.
The open markers represent the fit intercepts via different variables: $q_n^2$, $v_n$, and $v_n\Delta\delta$. The ensemble averages are also added with the solid markers in comparison. In general, the results for $\Delta\gamma_{112}\{\rm PP\}$ and $\Delta\gamma_{132}\{\rm PP\}$ qualitatively resemble those for $\Delta\gamma_{112}\{\rm RP\}$ and $\Delta\gamma_{132}\{\rm RP\}$, respectively. Therefore, the ESE method seems to work regardless of the event-plane type. We will further perform these analyses to the EBE-AVFD events in the following subsection to investigate whether the over-subtraction of background is model-dependent or a universal feature of this ESE approach.
The intercepts for $\Delta\gamma_{123}\{\rm PP\}$
are consistent with the ensemble average for the centrality range under study, indicating the failure of this ESE recipe for this observable.
This observation seems to echo the conclusion in Ref.~\cite{Subikash} that the underlying mechanism for $\Delta\gamma_{123}$ is different from that for the flow-related background in 
$\Delta\gamma_{112}$, and
thus $\Delta\gamma_{123}$ is not a good background estimate for $\Delta\gamma_{112}$.

It is remarkable that for all the aforementioned results, the intercepts via $q_n^2$ as the variable bear larger statistical uncertainties than those via the other two ($v_n$ and $v_n\Delta\delta$), though they are all consistent with each other.
This mostly results from the different fit functions to extract the intercepts. Therefore, $v_n$ and $v_n\Delta\delta$ are technically preferred over $q_n^2$ in projection of these $\Delta\gamma$ correlators to the zero-flow mode.

\subsection{EBE-AVFD}

The EBE-AVFD model~\cite{Shi:2017cpu,Jiang:2016wve,Shi:2019wzi} is a comprehensive simulation framework that describes the dynamical CME transport for quark currents in addition to the relativistically expanding viscous QGP fluid, and properly models major sources of background correlations, such as LCC and resonance decays. 

The initial conditions for entropy density ($s$) profiles and the initial electromagnetic field are fluctuated according to the event-by-event nucleon configuration in the Monte Carlo Glauber simulations~\cite{glauber}. The initial axial charge density ($n_5$) is introduced as being proportional to the corresponding local entropy density with a constant ratio. This ratio parameter can be varied to control the strength of the CME transport. For example, one can set $n_5/s$ to $0$, $0.1$, and $0.2$, to simulate scenarios of zero, modest, and strong CME signals, respectively. 

The hydrodynamic evolution is solved through two components: the bulk-matter collective flow and the dynamical CME transport. The former is managed by the VISH2+1 simulation package~\cite{Shen:2014vra}, which has been extensively tested and validated with relevant experimental data. The latter is described by anomalous hydrodynamic equations for the quark chiral currents on top of the bulk flow background. The magnetic-field-induced CME currents lead to a charge separation in the fireball. Additionally, the conventional transport processes like diffusion and relaxation for the quark currents are coherently included, and relevant details can be found in Refs.~\cite{Shi:2017cpu,Jiang:2016wve,Shi:2019wzi}. 

In the freeze-out process, the LCC effect is implemented by producing some charged hadron-antihadron pairs from the same fluid cell, with their momenta sampled independently in the local rest frame of the cell. In this study, a parameter of $P_\text{LCC} = 1/3$ is set to characterize the fraction of charged hadrons that are sampled in oppositely-charged pairs, while the rest of the hadrons are sampled independently. 
Finally, all the hadrons produced from the freeze-out hypersurface are further subject to hadron cascades through the UrQMD simulations~\cite{Bleicher:1999xi}, which account for various hadron resonance decay processes and automatically carry
their contributions to the charge-dependent correlations.

\subsubsection{Au+Au collisions at \texorpdfstring{$\sqrt{s_{\rm NN}} = 200$}{Lg} GeV}

\begin{figure}[h]
\centering
\includegraphics[width=0.39\textwidth]{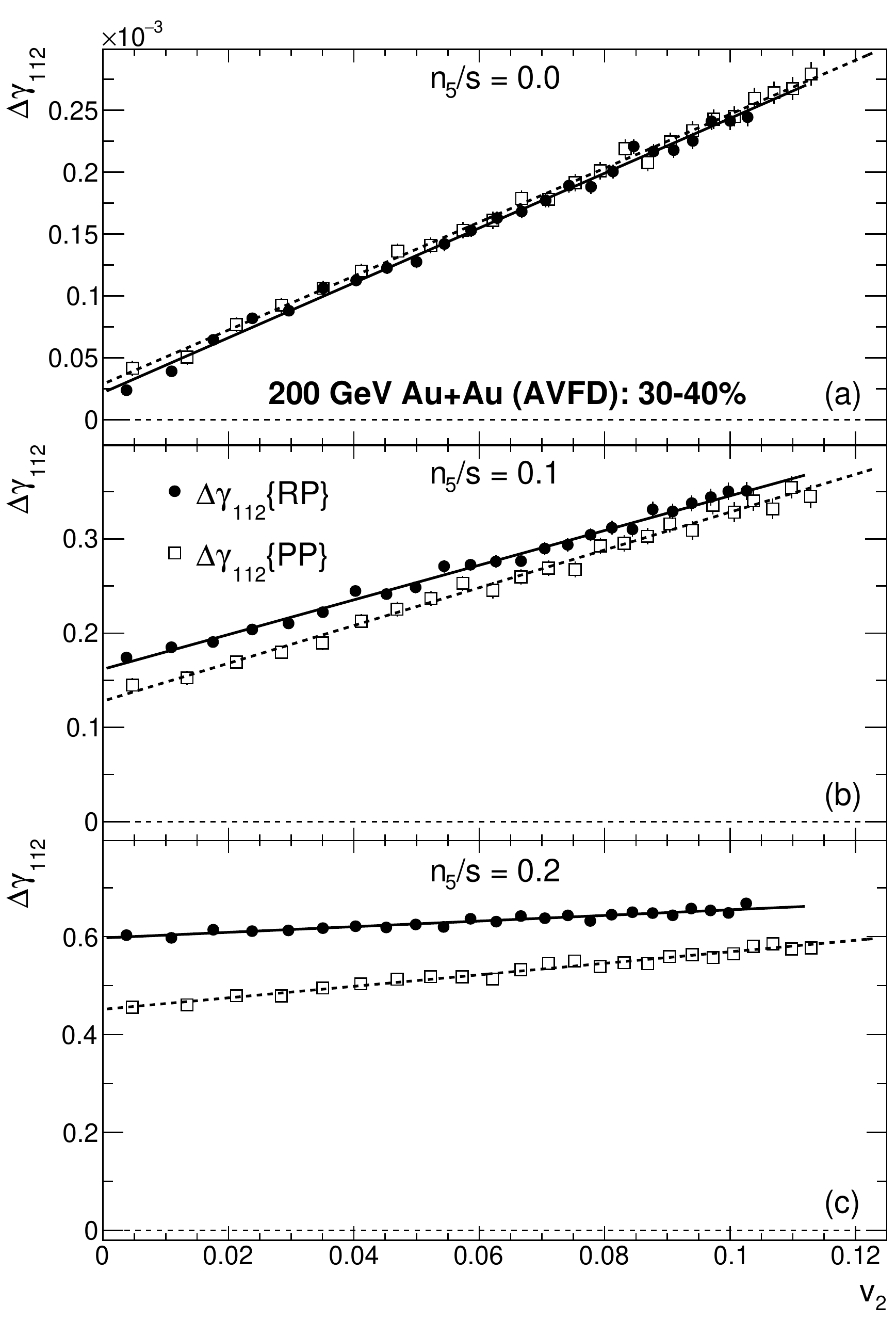}
\caption{EBE-AVFD calculations of $\Delta\gamma_{112}$ as a function of $v_{2}$ for $n_{5}/s$ of 0 (a), 0.1 (b), and 0.2 (c) in 30--40\% Au+Au collisions at $\sqrt{s_{\rm NN}}=200$ GeV. The results are fit with linear functions.}
\label{fig:AVFD_Au_Dg112_v2}
\end{figure} 

\begin{figure}[h]
\centering
\includegraphics[width=0.39\textwidth]{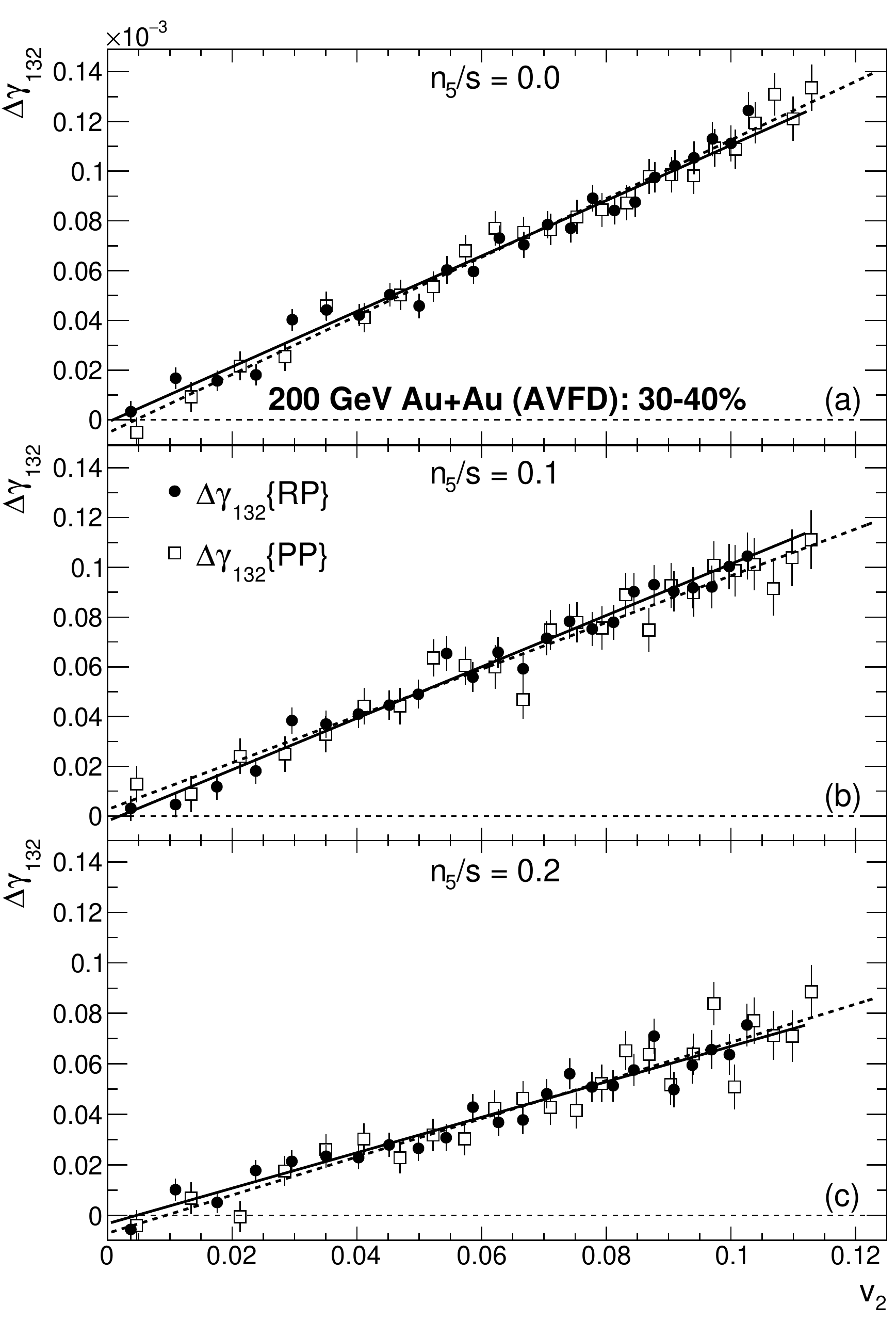}
\caption{EBE-AVFD simulations of $\Delta\gamma_{132}$ as a function of $v_{2}$ for $n_{5}/s$ of 0 (a), 0.1 (b), and 0.2 (c) in 30--40\% Au+Au collisions at $\sqrt{s_{\rm NN}}=200$ GeV. The results are fit with linear functions.}
\label{fig:AVFD_Au_Dg132_v2}
\end{figure} 

\begin{figure}[h]
\centering
\includegraphics[width=0.39\textwidth]{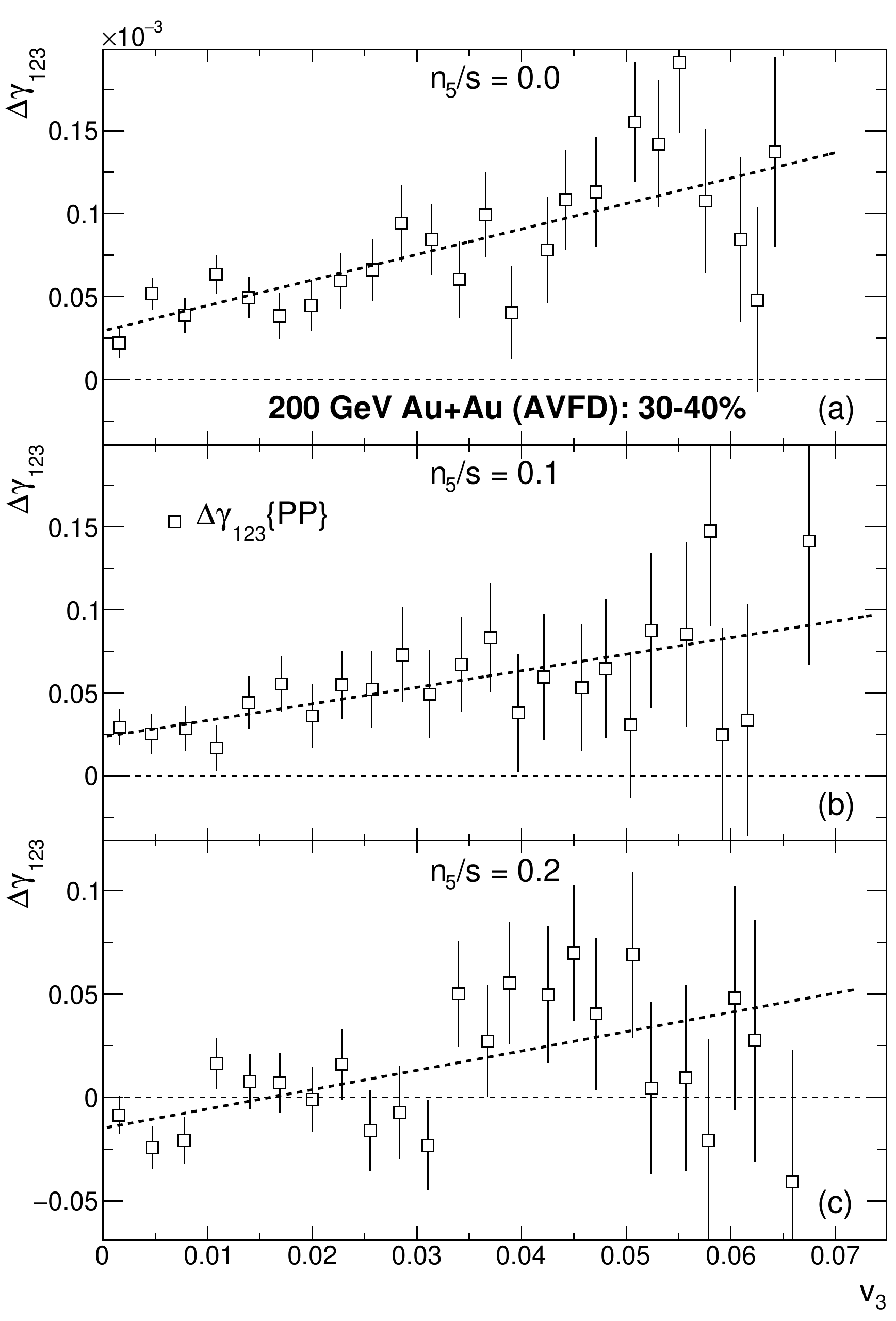}
\caption{EBE-AVFD calculations of $\Delta\gamma_{123}$ as a function of $v_{3}$ for $n_{5}/s$ of 0 (a), 0.1 (b), and 0.2 (c) in 30--40\% Au+Au collisions at $\sqrt{s_{\rm NN}}=200$ GeV. The results are fit with linear functions.}
\label{fig:AVFD_Au_Dg123_v3}
\end{figure} 

\begin{figure}[h]
\centering
\includegraphics[width=0.39\textwidth]{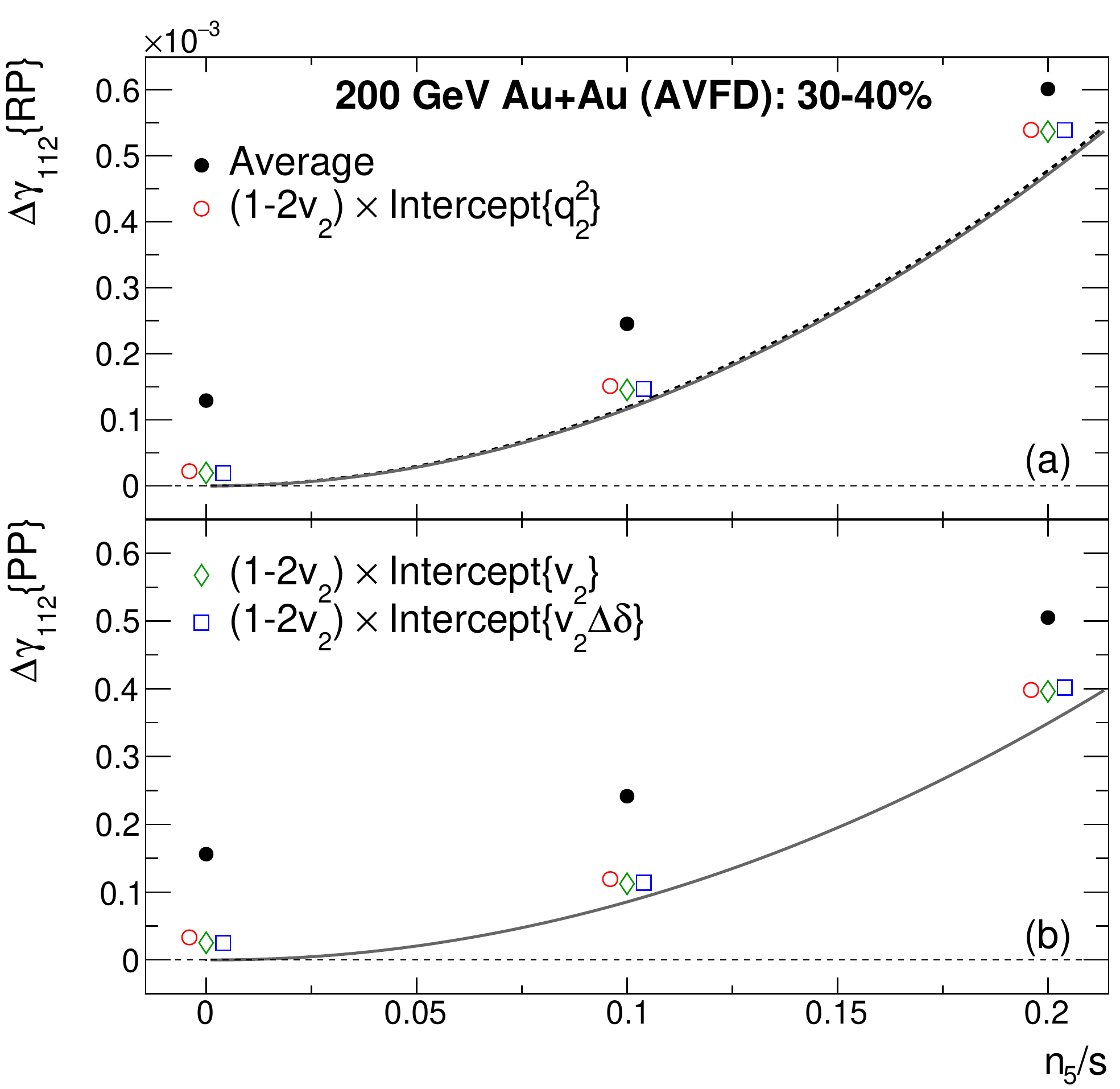}
\caption{$n_{5}/s$ dependence of $\Delta\gamma_{112}\{\rm RP\}$ (a) and $\Delta\gamma_{112}\{\rm PP\}$ (b) at the zero-flow mode for EBE-AVFD events of 30--40\% Au+Au collisions at $\sqrt{s_{\rm NN}}=200$ GeV. 
The open markers represent the fit intercepts via different variables: $q_2^2$, $v_2$, and $v_2\Delta\delta$. In comparison, the ensemble averages are also drawn with the solid markers. The solid and dashed lines are estimates for the CME signal, and are explained in the text.}
\label{fig:AVFD_Dg112_n5s}
\end{figure}

Figure~\ref{fig:AVFD_Au_Dg112_v2}
depicts the EBE-AVFD calculations of 
$\Delta\gamma_{112}\{\rm RP\}$ and $\Delta\gamma_{112}\{\rm PP\}$
as a function of $v_{2}$ for $n_{5}/s$ of 0 (a), 0.1 (b), and 0.2 (c) in 30--40\% Au+Au collisions at $\sqrt{s_{\rm NN}}=200$ GeV. The numbers of events are $9.6 \times10^7$, $5.9 \times 10^7$, and $7.7 \times 10^7$ for the cases of $n_5/s=0$, 0.1, and 0.2, respectively.
Note that the events are still binned with $q_2^2$ as done in the previous subsection. The solid (dashed) lines represent the linear fit functions to the results with respect to the RP (PP).
In the pure-background case ($n_5/s = 0$), the intercepts are positively finite, indicating that the flow-related background in the EBE-AVFD model cannot be completely removed by this ESE approach.
The magnitude of the intercept increases with increasing $n_5/s$, meeting the CME expectation.
At finite $n_5/s$ values, $\Delta\gamma_{112}\{\rm RP\}$ is above $\Delta\gamma_{112}\{\rm PP\}$, as the RP is more closely correlated with the magnetic-field direction, and hence $\Delta\gamma_{112}\{\rm RP\}$ contains a larger CME signal than $\Delta\gamma_{112}\{\rm PP\}$.
For simplicity, we do not show the similar results as a function of $q_2^2$ or $v_2\Delta\delta$, but the corresponding intercepts will be presented in Fig.~\ref{fig:AVFD_Dg112_n5s}.

\begin{figure}[bt]
\centering
\includegraphics[width=0.39\textwidth]{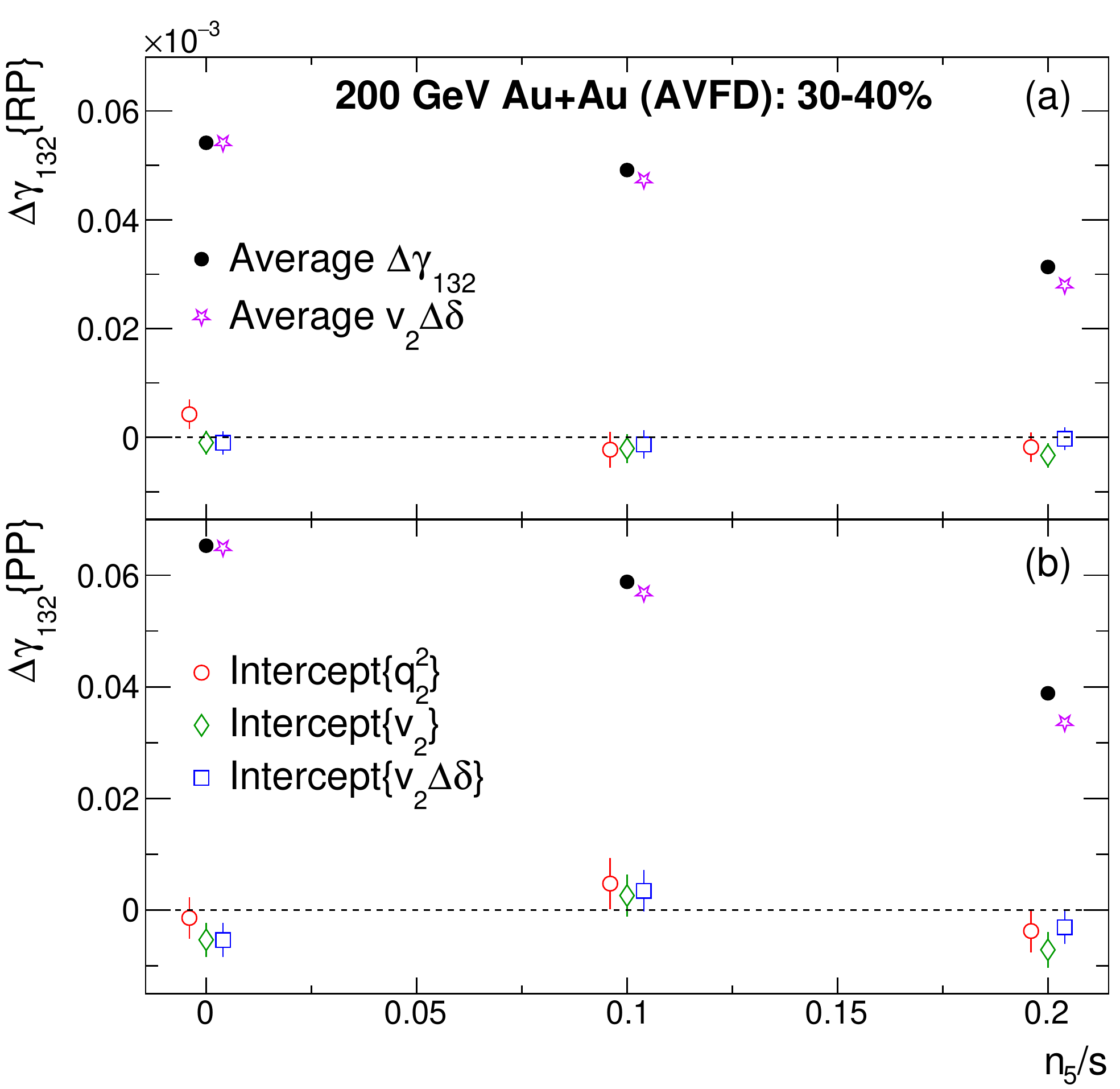}
\caption{$n_{5}/s$ dependence of $\Delta\gamma_{132}\{\rm PP\}$ (a) and $\Delta\gamma_{132}\{\rm PP\}$ (b) at the zero-flow mode for EBE-AVFD events of 30--40\% Au+Au collisions at $\sqrt{s_{\rm NN}}=200$ GeV.
The open markers represent the fit intercepts via different variables: $q_2^2$, $v_2$, and $v_2\Delta\delta$. In comparison, the ensemble averages for $\Delta\gamma_{132}$ and $v_2\Delta\delta$ are also drawn.}
\label{fig:AVFD_Dg132_n5s}
\end{figure}

Figure~\ref{fig:AVFD_Au_Dg132_v2}
delineates the EBE-AVFD simulations of 
$\Delta\gamma_{132}\{\rm RP\}$ and $\Delta\gamma_{132}\{\rm PP\}$
as a function of $v_{2}$ for $n_{5}/s$ of 0 (a), 0.1 (b), and 0.2 (c) in 30--40\% Au+Au collisions at $\sqrt{s_{\rm NN}}=200$ GeV. For all the $n_5/s$ values under study, the intercept of the linear fit is always consistent with zero. Similar intercept results extracted via $q_2^2$ or $v_2\Delta\delta$ will be summarized in Fig.~\ref{fig:AVFD_Dg132_n5s}.

\begin{figure}[h]
\centering
\includegraphics[width=0.39\textwidth]{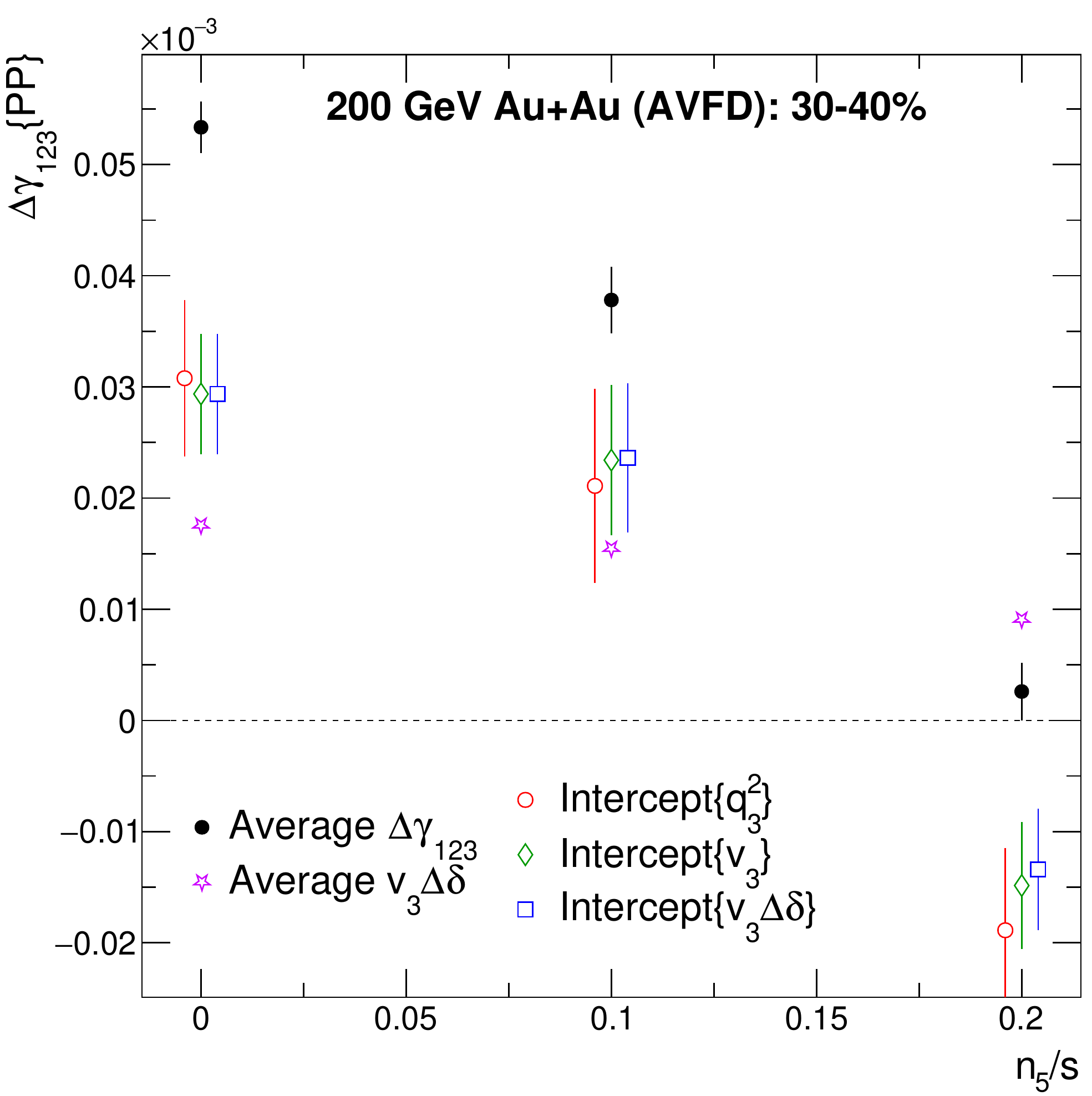}
\caption{$n_{5}/s$ dependence of $\Delta\gamma_{123}\{\rm PP\}$ at the zero-flow mode for EBE-AVFD events of 30--40\% Au+Au collisions at $\sqrt{s_{\rm NN}}=200$ GeV.
The open markers represent the fit intercepts via different variables: $q_3^2$, $v_3$, and $v_3\Delta\delta$. In comparison, the ensemble averages for $\Delta\gamma_{123}$ and $v_3\Delta\delta$ are also drawn.}
\label{fig:AVFD_Dg123_n5s}
\end{figure}

Figure~\ref{fig:AVFD_Au_Dg123_v3}
presents the EBE-AVFD calculations of 
 $\Delta\gamma_{123}\{\rm PP\}$
as a function of $v_{3}$ for $n_{5}/s$ of 0 (a), 0.1 (b), and 0.2 (c) in 30--40\% Au+Au collisions at $\sqrt{s_{\rm NN}}=200$ GeV. Note that the events are still binned with $q_3^2$ as done in the previous subsection. The intercept of the linear fit seems to decrease with increasing $n_5/s$, which will be further discussed in Fig.~\ref{fig:AVFD_Dg123_n5s}, together with
similar intercept results extracted via $q_3^2$ and $v_3\Delta\delta$.

We extract the ESE intercepts of $\Delta\gamma_{112}\{\rm RP\}$ and $\Delta\gamma_{112}\{\rm PP\}$ via $q_2^2$, $v_2$, and $v_2\Delta\delta$, for EBE-AVFD events of 30--40\% Au+Au collisions at $\sqrt{s_{\rm NN}} = 200$ GeV, and present the results corrected with $(1-2v_2)$ as a function of $n_5/s$ in Fig.~\ref{fig:AVFD_Dg112_n5s}. The conventional ensemble average values are also shown in comparison. In the pure-background scenario ($n_5/s=0$), although the ESE intercepts do not completely remove the residue background, they do suppress the background contribution roughly by a factor of 6 relative to the ensemble average for both $\Delta\gamma_{112}\{\rm RP\}$ and $\Delta\gamma_{112}\{\rm PP\}$.
In the cases of finite $n_5/s$ values, we estimate the CME contribution in two ways. As pointed out in Ref.~\cite{CME_methods}, with $\Psi_{\rm RP}$ known in the model, we can directly calculate $a_{1,\pm}$, and 
utilize the following relation to estimate the CME contribution in $\Delta\gamma_{112}\{\rm RP\}$:
\begin{eqnarray}
& & \Delta\gamma^{\rm CME}_{112}\{\rm RP\} \nonumber \\
&=& \Delta\gamma_{112}\{{\rm RP}\} - \Delta\gamma_{112}\{{\rm RP}\}|_{n_{5}/s=0}  \label{eq:Superposition1} \\
&=& (a_{1,+}^2 + a_{1,-}^2)/2 - a_{1,+}a_{1,-}.
\label{eq:Superposition2}
\end{eqnarray}

In Fig.~\ref{fig:AVFD_Dg112_n5s}(a), the solid line stands for a second-order polynomial fit to the quantity in Eq.~(\ref{eq:Superposition1}), whereas the dashed line denotes that to the quantity in Eq.~(\ref{eq:Superposition2}). The good consistency between the two estimates corroborates the relation in Eqs.~(\ref{eq:Superposition1}) and (\ref{eq:Superposition2}). In Fig.~\ref{fig:AVFD_Dg112_n5s}(b), where the PP is used in the analysis, only the solid line is drawn to represent $(\Delta\gamma_{112}\{{\rm PP}\} - \Delta\gamma_{112}\{{\rm PP}\}|_{n_{5}/s=0})$. In all the cases, the ESE results are much closer to the true CME signal than the ensemble average.

With the estimated $\Delta\gamma_{112}^{\rm CME}$ values, we can easily calculate the fraction of the CME signal, $f_{\rm CME}$, in the ensemble average of $\Delta\gamma_{112}$ as well as in the corrected ESE intercepts.
Table~\ref{tab:fCME_Au} lists the EBE-AVFD calculations of $f_{\rm CME}$ for different observables in 30--40\% Au+Au collisions at $\sqrt{s_{\rm NN}}=200$ GeV, for $n_5/s = 0.1$ and $0.2$. In general, $f_{\rm CME}$ increases with increasing $n_5/s$, as expected. The values of $f_{\rm CME}\{\rm RP\}$ for the ensemble averages are significantly larger than those of $f_{\rm CME}\{\rm PP\}$ as explained before: the smaller $v_2\{\rm RP\}$ values cause smaller flow-induced backgrounds, whereas the RP is more closely correlated with the magnetic-field direction, leading to larger CME signals. The difference between $f_{\rm CME}\{\rm RP\}$ and $f_{\rm CME}\{\rm PP\}$ is reduced for the ESE intercepts, since the background is largely suppressed.
At $n_5/s=0.2$ where the CME signal is very strong, $f_{\rm CME}$ could reach around 88\% for the ESE intercepts, and drop by 10--20\% for the ensemble average depending on whether the RP or the PP is used.
With weaker CME signals at $n_5/s=0.1$, the advantage of the ESE intercepts over the ensemble average becomes more prominent in $f_{\rm CME}$.
On the other hand, the disadvantage of the ESE approach is also clear: the statistical uncertainty is about 2--4 times larger than that of the ensemble average. 

\begin{table}[tb]
\caption{The fraction of the CME signal, $f_{\rm CME}$, for the ensemble average of $\Delta\gamma_{112}$ and the corrected ESE intercepts in EBE-AVFD events of 30--40\% Au+Au collisions at $\sqrt{s_{\rm NN}}=200$ GeV, for $n_5/s = 0.1$ and $0.2$.}
\begin{tabular}{ c || c| c |c |c}
\hline
 $n_5/s=0.1$ & Average & ESE\{$q_2^2$\} & ESE\{$v_2$\} & ESE\{$v_2\Delta\delta$\} \\
\hline 
$f_{\rm CME}\{{\rm RP}\}$ (\%)& 47.4$\pm$0.5 & 76.9$\pm$1.7 & 80.0$\pm$1.6 & 79.3$\pm$1.5 \\
\hline
$f_{\rm CME}\{{\rm PP}\}$ (\%)& 35.4$\pm$0.6 & 71.7$\pm$2.7 & 76.2$\pm$2.6 & 75.1$\pm$2.5 \\ 
\hline \hline
\end{tabular}

\begin{tabular}{ c || c| c |c |c}
\hline
 $n_5/s=0.2$ & Average & ESE\{$q_2^2$\} & ESE\{$v_2$\} & ESE\{$v_2\Delta\delta$\} \\
\hline 
$f_{\rm CME}\{{\rm RP}\}$ (\%)& 78.5$\pm$0.2 & 87.5$\pm$0.5 & 87.9$\pm$0.4 & 87.6$\pm$0.4 \\
\hline
$f_{\rm CME}\{{\rm PP}\}$ (\%)& 69.1$\pm$0.3 & 87.7$\pm$0.8 & 88.1$\pm$0.7 & 86.9$\pm$0.7 \\ 
\hline \hline
\end{tabular}

\label{tab:fCME_Au}

\end{table}

Figure~\ref{fig:AVFD_Dg132_n5s} shows the $n_{5}/s$ dependence of $\Delta\gamma_{132}\{\rm PP\}$ (a) and $\Delta\gamma_{132}\{\rm PP\}$ (b) at the zero-flow mode for EBE-AVFD events of 30--40\% Au+Au collisions at $\sqrt{s_{\rm NN}}=200$ GeV.
The ensemble average values for $\Delta\gamma_{132}$ and $v_2\Delta\delta$ are also drawn in comparison.
Unlike the case of $\Delta\gamma_{112}$, $\Delta\gamma_{132}$ seems to vanish with the ESE technique in most cases, supporting the idea that $\Delta\gamma_{132}$ is approximately equal to $v_2\Delta\delta$~\cite{Subikash}, and hence should disappear at the zero-flow mode.
The equivalence relation between $\Delta\gamma_{132}$ and $v_2\Delta\delta$
also explains why the ensemble average of $\Delta\gamma_{132}$ decreases with increasing $n_5/s$: $v_2$ is basically constant over $n_5/s$, and $\Delta\delta$ is expected to decrease with increasing $n_5/s$~\cite{Shi:2019wzi}. Therefore, in the real-data analyses, $\Delta\gamma_{132}$ can be used as a systematic check on how well the ESE approach works in terms of the background removal.

Figure~\ref{fig:AVFD_Dg123_n5s} shows the $n_{5}/s$ dependence of $\Delta\gamma_{123}\{\rm PP\}$ at the zero-flow mode for EBE-AVFD events of 30--40\% Au+Au collisions at $\sqrt{s_{\rm NN}}=200$ GeV, with the ensemble averages for $\Delta\gamma_{123}$ and $v_3\Delta\delta$ drawn in comparison. Both the ESE intercepts and the ensemble average for $\Delta\gamma_{123}$ have a stronger dependence on $n_5/s$ than $v_3\Delta\delta$.
The flow-related contributions in $\Delta\gamma_{123}$ seem to be reduced by the ESE technique, but they do not disappear as in the case of $\Delta\gamma_{132}$. Although a further investigation is needed to better understand the mechanism behind $\Delta\gamma_{123}$, we can draw a similar conclusion as in Ref.~\cite{Subikash} that $\Delta\gamma_{123}$ is not a proper background estimate for $\Delta\gamma_{112}$.

\subsubsection{Ru+Ru and Zr+Zr collisions at \texorpdfstring{$\sqrt{s_{\rm NN}} = 200$}{Lg} GeV}

\begin{figure}[h]
\centering
\includegraphics[width=0.39\textwidth]{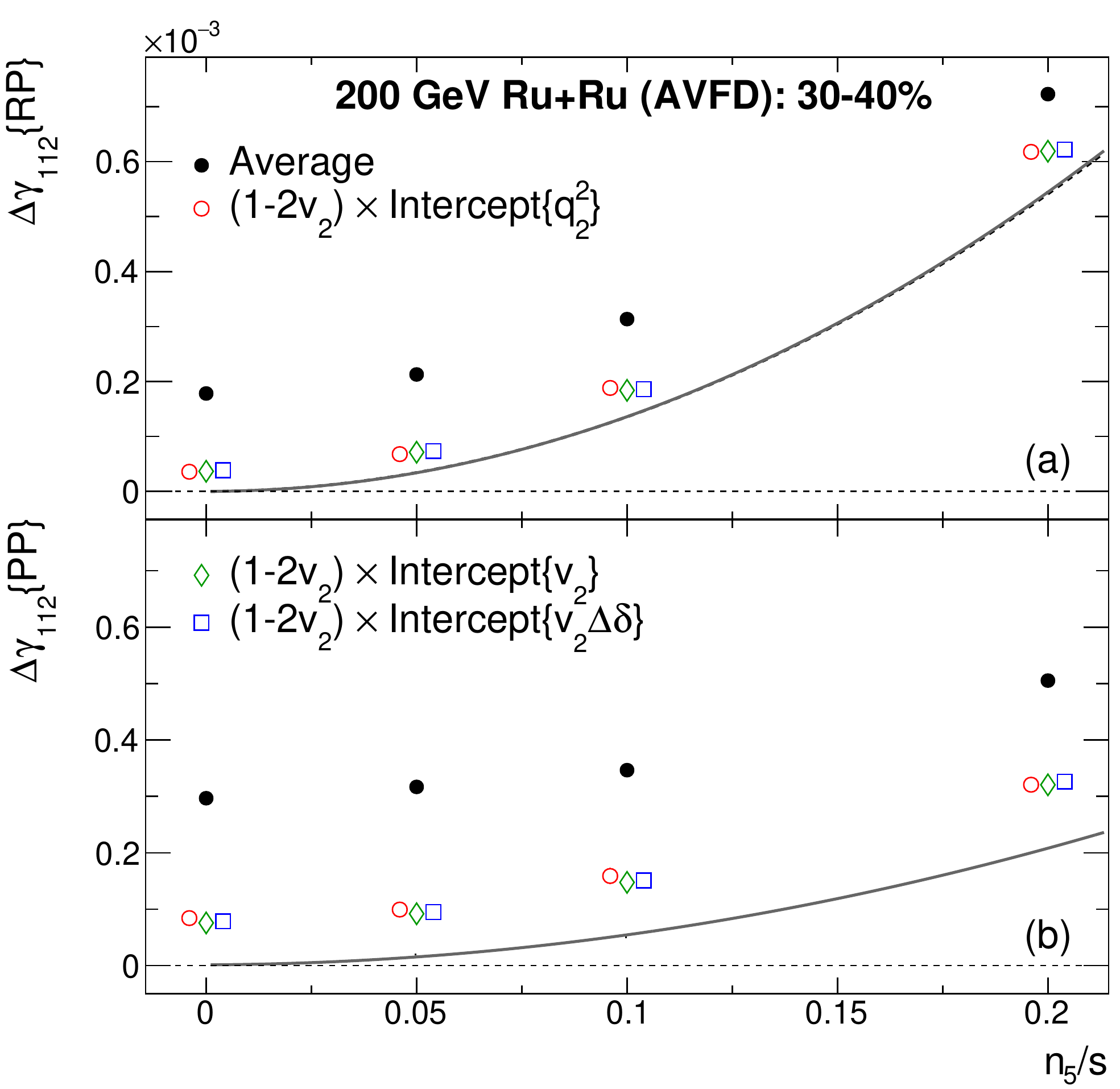}
\caption{$n_{5}/s$ dependence of $\Delta\gamma_{112}\{\rm RP\}$ (a) and $\Delta\gamma_{112}\{\rm PP\}$ (b) at the zero-flow mode for EBE-AVFD events of 30--40\% Ru+Ru collisions at $\sqrt{s_{\rm NN}}=200$ GeV. 
The open markers represent the fit intercepts via different variables: $q_2^2$, $v_2$, and $v_2\Delta\delta$. In comparison, the ensemble averages are also drawn with the solid markers. The solid and dashed lines are estimates for the CME signal, as explained in the text.}
\label{fig:AVFD_Ru_Dg112_n5s}
\end{figure} 

\begin{figure}[h]
\centering
\includegraphics[width=0.39\textwidth]{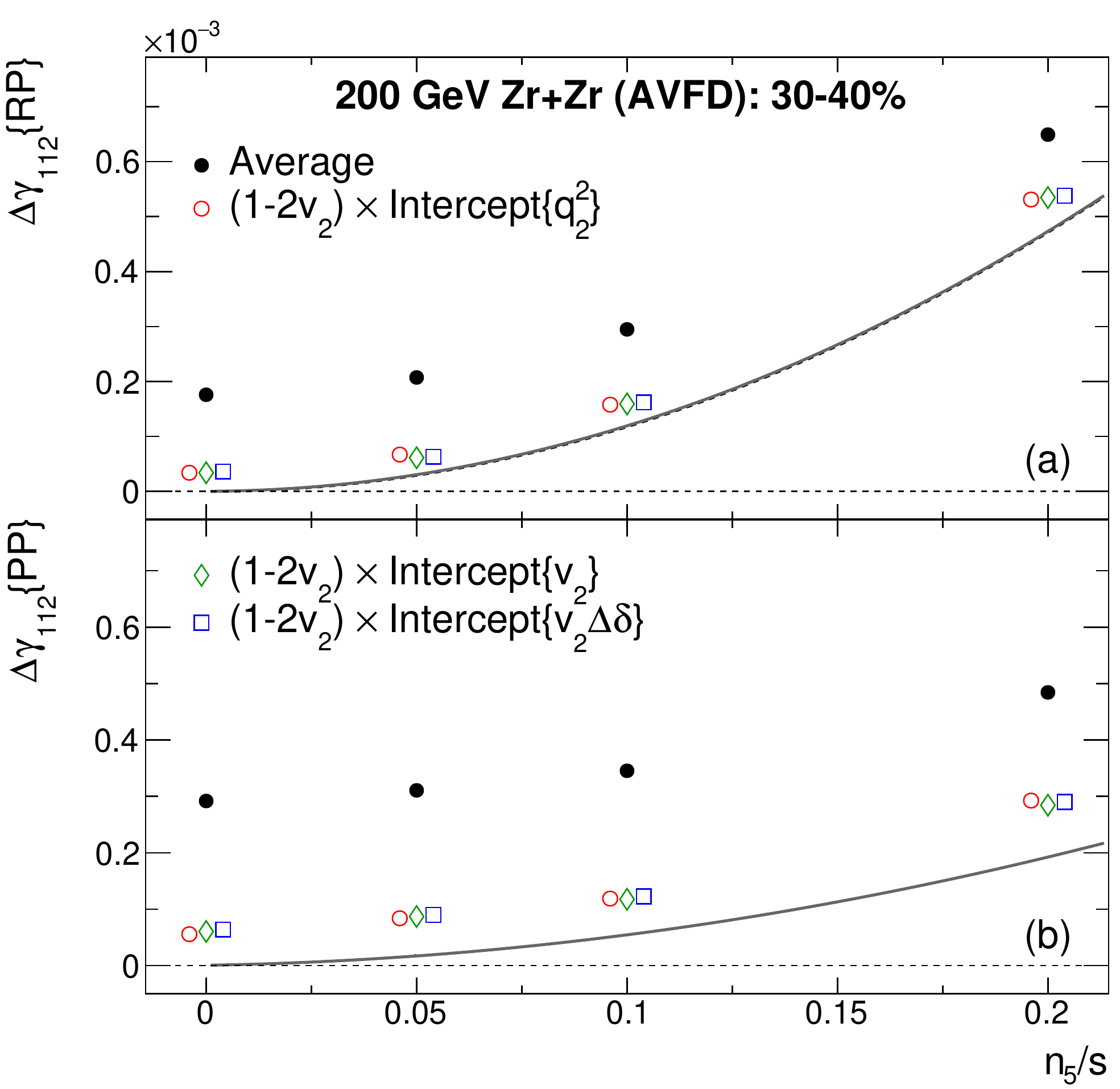}
\caption{$n_{5}/s$ dependence of $\Delta\gamma_{112}\{\rm RP\}$ (a) and $\Delta\gamma_{112}\{\rm PP\}$ (b) at the zero-flow mode for EBE-AVFD events of 30--40\% Zr+Zr collisions at $\sqrt{s_{\rm NN}}=200$ GeV. 
The open markers represent the fit intercepts via different variables: $q_2^2$, $v_2$, and $v_2\Delta\delta$. In comparison, the ensemble averages are also drawn with the solid markers. The solid and dashed lines are estimates for the CME signal, as explained in the text.}
\label{fig:AVFD_Zr_Dg112_n5s}
\end{figure} 

Recently the STAR Collaboration has completed the blind analysis of the isobar-collision data, without observing any predefined CME signature~\cite{STAR_isobar}. One possibility is that $f_{\rm CME}$ is much smaller in Ru+Ru and Zr+Zr than in Au+Au at the same $n_5/s$~\cite{Revisit_isobar}. We shall explore both the conventional ensemble average and the ESE intercepts for $\Delta\gamma_{112}$ along this direction with EBE-AVFD simulations. With our current precision, the results for Ru+Ru and Zr+Zr are consistent with each other for all the cases. The differentiation of the two isobaric systems requires a much larger event sample than what we currently use, and thus in this article we only focus on the common feature of the isobar collisions.

Figures~\ref{fig:AVFD_Ru_Dg112_n5s} and \ref{fig:AVFD_Zr_Dg112_n5s}
present the $n_{5}/s$ dependence of $\Delta\gamma_{112}\{\rm RP\}$ (a) and $\Delta\gamma_{112}\{\rm PP\}$ (b) at the zero-flow mode for EBE-AVFD events of 30--40\% Ru+Ru and Zr+Zr collisions, respectively, at $\sqrt{s_{\rm NN}}=200$ GeV. The numbers of events are $6.0 \ (4.8)\times10^7$, $3.7 \ (3.8)\times10^7$, $2.4 \ (7.1)\times10^7$, and $5.1 \ (5.6)\times10^7$ for Ru+Ru (Zr+Zr) collisions at $n_5/s=0$, 0.05, 0.1, and 0.2, respectively.
The ESE intercepts have been corrected with the factor of $(1-2v_2)$.
In comparison,
we also draw the ensemble averages, as well as the solid and dashed lines as estimates for the CME signal, obtained in the same way as previously done for Au+Au collisions. 
In the pure-background scenario ($n_5/s=0$), although the ESE intercepts do not completely remove the residue background, they do suppress the background contribution roughly by a factor of 5 relative to the ensemble average for both $\Delta\gamma_{112}\{\rm RP\}$ and $\Delta\gamma_{112}\{\rm PP\}$ in both isobaric systems. In the cases of finite $n_5/s$ values, the good consistency between the two estimates for the CME signal (solid and dashed lines) supports the relation in Eqs.~(\ref{eq:Superposition1}) and (\ref{eq:Superposition2}). In all the cases, the ESE results are much closer to the true CME signals than the ensemble averages.

\begin{table}[bt]
\caption{$f_{\rm CME}$ for the ensemble average of $\Delta\gamma_{112}$ and the corrected ESE intercepts in EBE-AVFD events of 30--40\% Ru+Ru collisions at $\sqrt{s_{\rm NN}}=200$ GeV, for $n_5/s = 0.05$, 0.1, and 0.2.}

\begin{tabular}{ c || c| c |c |c}
\hline
 $n_5/s=0.05$ & Average & ESE\{$q_2^2$\} & ESE\{$v_2$\} & ESE\{$v_2\Delta\delta$\} \\
\hline 
$f_{\rm CME}\{{\rm RP}\}$ (\%)& 16.3$\pm$1.7 & 51.0$\pm$6.7 & 48.5$\pm$5.8 & 47.2$\pm$5.5 \\
\hline
$f_{\rm CME}\{{\rm PP}\}$ (\%)& 6.3$\pm$2.1 & 20.2$\pm$7.1 & 21.8$\pm$7.5 & 21.1$\pm$7.3 \\ 
\hline \hline
\end{tabular}

\begin{tabular}{ c || c| c |c |c}
\hline
 $n_5/s=0.1$ & Average & ESE\{$q_2^2$\} & ESE\{$v_2$\} & ESE\{$v_2\Delta\delta$\} \\
\hline 
$f_{\rm CME}\{{\rm RP}\}$ (\%)& 43.2$\pm$1.4 & 71.9$\pm$3.5 & 73.6$\pm$3.1 & 72.7$\pm$3.1 \\
\hline
$f_{\rm CME}\{{\rm PP}\}$ (\%)& 14.4$\pm$2.2 & 31.3$\pm$5.7 & 33.7$\pm$5.9 & 33.0$\pm$5.7 \\ 
\hline \hline
\end{tabular}

\begin{tabular}{ c || c| c |c |c}
\hline
 $n_5/s=0.2$ & Average & ESE\{$q_2^2$\} & ESE\{$v_2$\} & ESE\{$v_2\Delta\delta$\} \\
\hline 
$f_{\rm CME}\{{\rm RP}\}$ (\%)& 75.3$\pm$0.5& 88.2$\pm$0.9 & 88.0$\pm$0.8 & 87.6$\pm$0.7 \\
\hline
$f_{\rm CME}\{{\rm PP}\}$ (\%)& 41.3$\pm$1.3 & 65.0$\pm$2.8 & 65.1$\pm$2.5 & 63.9$\pm$2.4 \\ 
\hline \hline
\end{tabular}

\label{tab:fCME_Ru}
\end{table}

\begin{table}[bt]
\caption{$f_{\rm CME}$ for the ensemble average of $\Delta\gamma_{112}$ and the corrected ESE intercepts in EBE-AVFD events of 30--40\% Zr+Zr collisions at $\sqrt{s_{\rm NN}}=200$ GeV, for $n_5/s = 0.05$, 0.1, and 0.2.}

\begin{tabular}{ c || c| c |c |c}
\hline
 $n_5/s=0.05$ & Average & ESE\{$q_2^2$\} & ESE\{$v_2$\} & ESE\{$v_2\Delta\delta$\} \\
\hline 
$f_{\rm CME}\{{\rm RP}\}$ (\%)& 15.2$\pm$1.7 & 46.9$\pm$6.7 & 51.3$\pm$7.0 & 49.8$\pm$6.7 \\
\hline
$f_{\rm CME}\{{\rm PP}\}$ (\%)& 6.1$\pm$2.2 & 22.5$\pm$8.7 & 21.7$\pm$8.2 & 21.0$\pm$7.9 \\ 
\hline \hline
\end{tabular}

\begin{tabular}{ c || c| c |c |c}
\hline
 $n_5/s=0.1$ & Average & ESE\{$q_2^2$\} & ESE\{$v_2$\} & ESE\{$v_2\Delta\delta$\} \\
\hline 
$f_{\rm CME}\{{\rm RP}\}$ (\%)& 40.3$\pm$1.1 & 75.4$\pm$2.8 & 74.7$\pm$2.5 & 73.3$\pm$2.4 \\
\hline
$f_{\rm CME}\{{\rm PP}\}$ (\%)& 15.5$\pm$1.7 & 45.1$\pm$5.9 & 45.6$\pm$5.7 & 43.7$\pm$5.4 \\ 
\hline \hline
\end{tabular}

\begin{tabular}{ c || c| c |c |c}
\hline
 $n_5/s=0.2$ & Average & ESE\{$q_2^2$\} & ESE\{$v_2$\} & ESE\{$v_2\Delta\delta$\} \\
\hline 
$f_{\rm CME}\{{\rm RP}\}$ (\%)& 72.9$\pm$0.6 & 89.1$\pm$1.0 & 88.5$\pm$0.9 & 88.0$\pm$0.9 \\
\hline
$f_{\rm CME}\{{\rm PP}\}$ (\%)& 39.8$\pm$1.3 & 65.8$\pm$3.1 & 67.7$\pm$2.9 & 66.4$\pm$2.8 \\ 
\hline \hline
\end{tabular}

\label{tab:fCME_Zr}

\end{table}

Tables~\ref{tab:fCME_Ru} and \ref{tab:fCME_Zr}
list the EBE-AVFD calculations of $f_{\rm CME}$ for different observables in 30--40\% Ru+Ru and Zr+Zr collisions, respectively, at $\sqrt{s_{\rm NN}}=200$ GeV, for $n_5/s = 0.05$, $0.1$, and $0.2$.
Compared with Au+Au collisions in the same centrality range,
the isobar collisions produce weaker magnetic fields. But the dilution effect in the $\gamma_{112}$ correlation is also weaker in the smaller systems~\cite{dilution}. As a result, $\Delta\gamma_{112}^{\rm CME}\{\rm RP\}$ remains almost the same from Au+Au to the isobar collisions, as shown with the similar curves in the upper panels of Figs.~\ref{fig:AVFD_Dg112_n5s}, \ref{fig:AVFD_Ru_Dg112_n5s} and \ref{fig:AVFD_Zr_Dg112_n5s}. On the other hand, the background correlation is also less diluted in the isobar collisions, making 
the $f_{\rm CME}\{\rm RP\}$ values for both the ensemble average and the ESE intercepts slightly lower than the Au+Au results in the same centrality range at the same $n_5/s$.
When the PP is used instead of the RP,
$f_{\rm CME}\{\rm PP\}$ for the ensemble average of $\Delta\gamma_{112}$ is much lower in the isobar collisions than that in Au+Au, consistent with the findings in Ref~\cite{Revisit_isobar}, because the smaller-system isobaric systems involve a larger fluctuation effect, which not only reduces the CME signal, but also increases the flow-related background, compared with the results using the RP. In this aspect, the ESE intercepts show a better performance than the ensemble average, with the background largely suppressed.

In addition to the expectation of $f_{\rm CME}$ obtained with different methods, these tables also provide the estimate of the statistical significance of the observables under study. 
We use the ratio of the mean value to its statistical uncertainty to quantify the statistical significance for $f_{\rm CME}$.
The ESE approach renders better CME signal fractions than the ensemble average, but lower statistical significance values, especially for $n_5/s \ge 0.1$.
When the CME signal is weak, e.g., with $n_5/s = 0.05$, similar significance levels are reached for the ensemble average and the ESE intercepts. If the CME signal is even weaker, we expect the ESE technique to surpass the ensemble average at the statistical significance. 

\section{Summary and Discussion}

We have presented a method study of the ESE technique that suppresses the flow-related background in the CME-searching observable, $\gamma_{112}$.
The previous work~\cite{ESE1} is extended in several directions. First, while sticking to the event-shape handle, $q_n^2$, we achieve the zero-flow mode via three variables: $q_n^2$, $v_n$, and $v_n\Delta\delta$. The fit intercepts thus obtained are consistent with each other, but the results using $v_n$ and $v_n\Delta\delta$ yield similarly smaller statistical uncertainties.

Second, we have examined the responses of both $\gamma_{112}$ and its variations ($\gamma_{132}$ and $\gamma_{123}$) to the change in the event shape. We will save $\gamma_{112}$ for later discussions. In all the cases of model simulations, $\Delta\gamma_{132}$ behaves like $v_2\Delta\delta$, and therefore always vanish at the zero-flow mode. This general feature of $\Delta\gamma_{132}$ can be used as a sanity check for real-data analyses. On the other hand, the behavior of $\Delta\gamma_{123}$ is model-dependent.
The pure-background AMPT calculations show that $\Delta\gamma_{123}$ is independent of the event shape selection, while 
EBE-AVFD simulations reveal that the magnitude of $\Delta\gamma_{123}$ does decrease towards the zero-flow mode.
However, neither the ensemble average of $\Delta\gamma_{123}$ nor the ESE intercepts can be explained by the flow driven mechanism. We conclude that $\Delta\gamma_{123}$ is formed differently from the flow-induced background in $\Delta\gamma_{112}$, and $\Delta\gamma_{123}$ does not represent a good background estimate for the $\Delta\gamma_{112}$ measurements.

Third, the EBE-AVFD model not only corroborates AMPT in the pure-background scenario, but also tests the sensitivity of the ESE recipe to the CME signal. When there is no CME input, the ESE intercepts of $\Delta\gamma_{112}$ are consistent with zero or slightly negative in AMPT events, and slightly positive in EBE-AVFD. Thus the residue background in the ESE intercepts is model-dependent, probably relying on the implementation details of the flowing resonances that decay into oppositely-charged particles. But the bottom line is that the ESE technique suppresses the flow-related background at least by a factor of 6 relative to the ensemble average in 30--40\% Au+Au collisions at $\sqrt{s_{\rm NN}} = 200$ GeV.
In the same collision system, when a finite $n_5/s$ of 0.1 or 0.2 is applied, the CME fraction, $f_{\rm CME}$, is substantially higher for the ESE intercepts than that for the ensemble average of $\Delta\gamma_{112}$.

Fourth, we have also explored the EBE-AVFD events of 30--40\% isobar collisions
(Ru+Ru and Zr+Zr) at $\sqrt{s_{\rm NN}} = 200$ GeV.
Compared with the larger Au+Au collision system in the same centrality range at the same $n_5/s$, $f_{\rm CME}$ for the ensemble average of $\Delta\gamma_{112}$ is largely reduced in the isobar collisions when the participant plane is used, but that for the ESE intercepts under the same condition only becomes slightly lower. According to the EBE-AVFD estimation, we expect the ESE method to surpass the ensemble average of $\Delta\gamma_{112}$ in the statistical significance of the CME signal, when $n_5/s$ is smaller than 0.05. This point bears a realistic importance, since the recently-posted STAR data~\cite{STAR_isobar} imply that $f_{\rm CME}$ for the ensemble average of $\Delta\gamma_{112}$ may be very small in such isobar collisions. Although the prospect of discovering the CME with the isobar data is waning, the alternative measurements using the ESE approach and the spectator plane (as a proxy for the RP) provide a more promising opportunity for future analyses over other methods like the ensemble average using the PP.


\begin{acknowledgments}
{The authors thank Zi-Wei Lin and Guo-Liang Ma for providing the AMPT code, and Yufu Li for generating the EBE-AVFD events. We are especially grateful to Aihong Tang and Sergei Voloshin for the fruitful discussions. R. Milton, G. Wang, M. Sergeeva, and H. Huang are supported
by the U.S. Department of Energy under Grant No. DE-FG02-88ER40424. J. Liao is supported by the NSF Grant No. PHY-1913729 and the U.S. Department of Energy, Office of Science, Office of Nuclear Physics, within the framework of the Beam Energy Scan Theory (BEST) Topical Collaboration. S. Shi is supported by the Natural Sciences and Engineering Research Council of Canada and the Fonds de recherche du Qu\'ebec - Nature et technologies (FRQNT) through the Programmede Bourses d'Excellencepour \'Etudiants \'Etrangers (PBEEE) scholarship.
}

\end{acknowledgments}
\nocite{*}
{}

\end{document}